\documentclass[preprint,12pt]{elsarticle}

\usepackage{graphicx}
\usepackage{amsmath,amssymb}
\usepackage{bm}
\usepackage{hyperref}
\newcommand{\FloatBarrier}{\par\medskip}
\hypersetup{colorlinks=true,linkcolor=blue,citecolor=blue,urlcolor=blue}
\DeclareUnicodeCharacter{2219}{\ensuremath{\cdot}}
\DeclareUnicodeCharacter{21CC}{\ensuremath{\rightleftharpoons}}
\DeclareUnicodeCharacter{03B2}{\ensuremath{\beta}}
\DeclareUnicodeCharacter{2261}{\ensuremath{\equiv}}
\DeclareUnicodeCharacter{2208}{\ensuremath{\in}}
\DeclareUnicodeCharacter{2026}{\ensuremath{\ldots}}
\DeclareUnicodeCharacter{202F}{\,}

\journal{Journal of Computational Physics}

\begin{document}

\begin{frontmatter}

\title{A conservative coupling method of sharp-interface and multi-species model for compressible reacting gas-liquid flows with phase change}

\author[aff1,aff2]{Jiaxi Song}
\author[aff1,aff2]{Yunzhang Tian}
\author[aff1,aff2,aff3]{Shucheng Pan\corref{cor1}}
\cortext[cor1]{Corresponding author. Email address: shucheng.pan@nwpu.edu.cn}

\address[aff1]{School of Aeronautics, Northwestern Polytechnical University, Xi'an 710072, China}
\address[aff2]{National Key Laboratory of Aircraft Configuration Design, Xi'an 710072, China}
\address[aff3]{Institute of Extreme Mechanics, Northwestern Polytechnical University, Xi'an 710072, China}

\begin{abstract}
In this paper, a conservative sharp-interface and diffuse-interface coupling method is developed for
compressible two-phase multi-species flows with phase change and chemical reactions. The
liquid--gas interface is represented by a sharp-interface model, whereas a diffuse-interface
model treats the transport and chemical reactions of gas-phase species.
Conservation is enforced by coupling the two phases through interfacial fluxes obtained from a
multi-species phase-change Riemann problem. The original single-species four-wave Riemann solver
is extended to multi-species gas mixtures by modifying the interfacial energy jump condition.
Interfacial mass transfer is restricted to the condensable vapor species. Accordingly, both the
interfacial energy jump condition and the gas-mixture energy-exchange flux are constructed using
the internal energy of the phase-changing vapor species rather than the mixture internal energy.
With this species-selective energy coupling, an approximate multi-species Riemann solver is
constructed that retains the four-wave structure while avoiding the multidimensional nonlinear
root-finding required by the exact solution. A series of numerical tests,
including impulsive evaporation and condensation, reacting aluminum vaporization,
shock-droplet interaction, and detonation-droplet interaction, are performed to assess the
accuracy and robustness of the method.
The numerical results agree well with
reference solutions and benchmark data, demonstrating that the present method resolves the
effects of phase change and chemical reactions in compressible multi-species multiphase flows
while preserving conservative interfacial coupling.
\end{abstract}


\end{frontmatter}

\section{Introduction}

Multiphase flows with phase change arise in a wide range of scientific and industrial
applications, including cavitation
\cite{bibal2024compressibleCavitation,mossier2025gprPhaseTransition}, scramjet engines
\cite{powell2001hytechScramjet}, aluminized explosives
\cite{sielicki2022aluminiumPowderExplosions}, and liquid-fueled detonation-based propulsion
\cite{wolanski2013detonativePropulsion,lu2014rotatingDetonation,zhu2024liquidFuelsRDE}. In many
of these strongly compressible problems, such as shock-droplet or detonation-droplet
interactions, phase change is coupled with rapid pressure-wave propagation, species transport,
and finite-rate chemistry
\cite{das2020shockVaporizationDroplets,tarey2024reactiveDroplet,xu2024detonationWaterDroplet,
huang2024evaporatingDropletShock,musick2023heterogeneousDetonations}.

Numerical approaches developed for such compressible multiphase-flow problems have generally
followed two routes: diffuse-interface methods and sharp-interface methods. In diffuse-interface
methods, the material interface is spread over a finite number of grid cells and the
interfacial region is described by mixture variables \cite{adebayo2025diffuseReview}. A
mixture model or an equation of state is therefore required to define thermodynamically
consistent pressure, temperature, composition, and phase state in mixed cells
\cite{allaire2002fiveEquation,saurel2009relaxation}. Recent work has extended diffuse-interface
methods to phase-changing flows by coupling pressure-based mixture equations with a
phase-indicator transport equation, and a related formulation has been developed for
low-Mach-number multicomponent two-phase flows
\cite{demou2022diffuseMassTransfer,salimi2025lowMachMulticomponent}. These approaches incorporate
phase-change mass transfer and species transport within the diffuse interfacial description
without explicitly reconstructing a sharp boundary. In sharp-interface methods, the material
interface is
represented as a discontinuity in the flow field. Different approaches have been developed to
describe or track the interface with a non-smeared representation, including front-tracking
methods \cite{unverdi1992frontTracking}, volume-of-fluid methods
\cite{hirt1981vof,rider1998volumeTracking}, level-set methods \cite{osher1988levelSet}, and
immersed-interface methods \cite{leveque1994immersedInterface,lee2003immersedInterface}. In the
present work, the interface is represented by a level-set function, where the liquid--gas interface is
defined as the zero contour of a signed-distance function. This representation provides a
convenient way to compute interface normals and curvatures and to handle large interface
deformations without explicit interface reconstruction. Once the interface geometry is
available, the next issue is how phase change is coupled to the sharp interface. Sharp-interface
methods with phase change have been extensively studied for incompressible flows
\cite{gibou2007levelSetPhaseChange,pang2023conservativeSharpInterface,
dodd2025vofPressureCorrection} and low-Mach-number flows
\cite{tanguy2007vaporizing,rueda2016ghostFluidBoiling,cipriano2024multicomponentDroplet}.
Front-tracking, level-set, ghost-fluid, and volume-of-fluid methods have been applied to boiling
flows \cite{juric1998boiling,rueda2016ghostFluidBoiling}, evaporation and Stefan-flow problems
\cite{gibou2007levelSetPhaseChange,tanguy2007vaporizing}, and droplet vaporization
\cite{tryggvason2005dnsPhaseChange,cipriano2024multicomponentDroplet}. Recent work has further
clarified consistency, interface velocity, and conservation issues in sharp-interface or
interface-resolved phase-change simulations
\cite{boniou2022consistency,lu2023velocityDecomposition,pang2023conservativeSharpInterface,
dodd2025vofPressureCorrection,cipriano2024multicomponentDroplet}. These studies demonstrate the
importance of accurately treating the interface motion and the mass-transfer-induced velocity
jump. Most of them, however, are not designed for strong compressibility, shock-interface
interaction, or chemically reacting multi-species gases.

Using sharp-interface methods, several approaches have been developed for the numerical
simulation of compressible multiphase flows with phase change using level-set interface tracking
and ghost-fluid treatments
\cite{houim2013ghostFluidPhaseChange,das2020shockVaporizationDroplets,
das2021reactingAluminumDroplets}. Houim and Kuo~\cite{houim2013ghostFluidPhaseChange} pioneered
the extension of the GFM~\cite{fedkiw1999ghostFluid} to compressible reacting flows with phase
change by solving a modified interfacial Riemann problem that accounts for phase change and
surface tension. Das and
Udaykumar~\cite{das2020shockVaporizationDroplets,das2021reactingAluminumDroplets} subsequently
simplified the implementation of the interfacial jump conditions and avoided the numerical
instability associated with rotating the deviatoric stress tensor in
Ref.~\cite{houim2013ghostFluidPhaseChange} by rotating the velocity field instead. A more complete
and self-consistent local interface closure was developed by Fechter
et al.~\cite{fechter2017sharpInterfacePhaseTransition,fechter2018riemannPhaseTransition}, who
formulated a general phase-change Riemann problem and an approximate solver for compressible
liquid-vapor flow with phase change and surface tension. Their method includes the latent-heat
energy jump and provides a complete set of local jump conditions for pure liquid-vapor systems.
Following a different modeling route, J\"ons and
Munz~\cite{jons2023riemannPhaseTransition} constructed exact and approximate two-phase Riemann
solvers based on the Euler-Fourier system, with the interfacial mass-transfer rate closed through
classical irreversible thermodynamics and Onsager coefficients rather than a latent-heat jump
relation. These local Riemann-problem closures
\cite{fechter2017sharpInterfacePhaseTransition,fechter2018riemannPhaseTransition,
jons2023riemannPhaseTransition}, like the preceding GFM-based methods, are coupled to the bulk flow
through ghost states rather than strictly conservative interfacial fluxes. Consequently, mass,
momentum, and total energy are not exactly conserved across the interface in the discrete sense.
To address this limitation, Long et al.~\cite{long2023conservativePhaseChange} built on the
phase-change Riemann problem of Fechter et
al.~\cite{fechter2017sharpInterfacePhaseTransition,fechter2018riemannPhaseTransition} and
developed a conservative sharp-interface method based on interfacial flux coupling. The
interfacial fluxes are obtained from a four-wave approximate Riemann solver, avoiding the
multidimensional root-finding required by the exact solver. They also showed that the
phase-change mass flux should be evaluated from the states adjacent to the phase interface to
ensure numerical consistency. Separately, Wenzel and
Arienti~\cite{wenzel2023compressibleFramework} developed a conservative, interface-resolved
evaporation method based on an asymptotically preserving method applicable to both
compressible and incompressible regimes. Their phase-change verification and applications focus
on evaporation configurations rather than shock-driven phase change or condensation. Extending
such conservative phase-change coupling to a reacting multi-species gas introduces an additional
difficulty: only the condensable vapor crosses the liquid--gas interface, so the mass and energy
exchanges are species-selective and must be consistent with gas-phase species transport and
thermodynamics \cite{larrouturou1989multicomponentGas,johnsen2006wenoMulticomponent}.
Conservative interface treatments are available either for two-material reacting gases without
phase change \cite{xu2022interfaceTreatment} or for mainly incompressible or low-Mach-number
multicomponent evaporation \cite{cipriano2024multicomponentDroplet,salimi2025lowMachMulticomponent}.
A conservative sharp-interface flux formulation for strongly compressible, reacting multi-species
flows remains unavailable.

In this paper, we develop a conservative sharp-interface and diffuse-interface coupling method in which the
liquid--gas interface is represented by a sharp-interface model, whereas a diffuse-interface
formulation describes gas-phase species transport and chemical reactions. The conservative sharp-interface formulation is based on
Refs.~\cite{hu2006conservativeInterface,han2014adaptiveMultiresolution} and their extension to
phase change~\cite{long2023conservativePhaseChange}. On this basis, the single-species four-wave
phase-change Riemann solver of Long et
al.~\cite{long2023conservativePhaseChange} is extended to multi-species gas mixtures through a
modified interfacial energy jump condition. In the present method,
unlike the GFM-based methods of
Refs.~\cite{houim2013ghostFluidPhaseChange,das2020shockVaporizationDroplets,
das2021reactingAluminumDroplets},
conservation is ensured by implementing the coupling between the two phases through
interfacial fluxes. In
contrast to conservative sharp-interface methods
developed for pure liquid-vapor systems
\cite{lauer2012cavitationBubbleDynamics,paula2019waterDropExplosion,
long2023conservativePhaseChange}, the present method couples a liquid phase to a chemically
reacting multi-species gas containing both condensable vapor and non-condensable species, and
incorporates species diffusion and finite-rate chemical reactions.
The remainder of this paper is organized as follows. Section~\ref{sec:framework} presents the
conservative sharp-interface and diffuse-interface coupling method. Section~\ref{sec:riemann-problem} introduces the
multi-species phase-change Riemann problem and its approximate solver. Section~\ref{sec:results}
validates the method and examines evaporation, condensation, shock interaction with vaporizing,
condensing, and reacting droplets, and detonation-droplet interaction. Conclusions are given in
Section~\ref{sec:conclusions}.

\section{Conservative sharp-interface and diffuse-interface coupling method}
\label{sec:framework}

\subsection{Governing equations}

We consider a two-phase compressible system composed of a pure liquid and a multi-species gas
mixture. The entire flow domain $\Omega$ is
divided into two sub-domains, a
gas domain $\Omega_{\mathrm{gas}}$ and a liquid domain $\Omega_{\mathrm{liq}}$, separated by an
evolving interface $\Gamma(t)$. Each phase can be described by the Navier--Stokes equations

\begin{equation}
\frac{\partial \bm{U}}{\partial t}+\nabla\cdot\bm{F}=\nabla\cdot\bm{F}_v+\bm{S},\label{eq:eq1}
\end{equation}

\noindent where $\bm{U}$ is the vector of conservative variables, $\bm{F}$ and $\bm{F}_v$ denote
the convective and viscous fluxes, respectively. $\bm{S}$ is the source term accounting for
chemical reactions and surface tension.

The gas mixture comprises $N$ species, including the vapor species indexed by $k=1$
that can undergo phase change with the liquid phase and $N-1$ non-condensable species, and is governed by a
compressible four-equation model \cite{yi2019multicomponentFourEquation}. The liquid phase is
treated as a pure fluid. In the absence of species diffusion and chemical reactions, its governing
equations reduce to the standard single-component compressible Navier--Stokes equations
\cite{houim2013ghostFluidPhaseChange}. For the multi-species gas phase, the vectors in 
Eq.~\eqref{eq:eq1} are given by

\begin{equation}
\begin{gathered}
\begin{aligned}
\bm{U}
&=
\begin{bmatrix}
\rho\\
\rho u\\
\rho v\\
\rho w\\
\rho E\\
\rho Y_1\\
\vdots\\
\rho Y_{N-1}
\end{bmatrix},
\qquad
\bm{S}
&=
\begin{bmatrix}
0\\
\sigma\kappa\delta_\Gamma n_x\\
\sigma\kappa\delta_\Gamma n_y\\
\sigma\kappa\delta_\Gamma n_z\\
\sigma\kappa\delta_\Gamma\bm{u}\cdot\bm{n}\\
\dot{\omega}_1\\
\vdots\\
\dot{\omega}_{N-1}
\end{bmatrix},
\end{aligned}
\\[1em]
\begin{aligned}
\bm{F}(\bm{U})
&=
\begin{bmatrix}
\rho u & \rho v & \rho w\\
\rho u^2+p & \rho u v & \rho u w\\
\rho u v & \rho v^2+p & \rho v w\\
\rho u w & \rho v w & \rho w^2+p\\
(\rho E+p)u & (\rho E+p)v & (\rho E+p)w\\
\rho uY_1 & \rho vY_1 & \rho wY_1\\
\vdots & \vdots & \vdots\\
\rho uY_{N-1} & \rho vY_{N-1} & \rho wY_{N-1}
\end{bmatrix}.
\end{aligned}
\\[1em]
\begin{gathered}
\bm{F}_v(\bm{U})=
\\[-0.2em]
\resizebox{0.90\linewidth}{!}{$\displaystyle
\left[
\begin{array}{@{\hspace{0.25em}}c@{\hspace{0.55em}}c@{\hspace{0.55em}}c@{\hspace{0.25em}}}
0 & 0 & 0\\
\tau_{xx} & \tau_{yx} & \tau_{zx}\\
\tau_{xy} & \tau_{yy} & \tau_{zy}\\
\tau_{xz} & \tau_{yz} & \tau_{zz}\\
u\tau_{xx}+v\tau_{xy}+w\tau_{xz}-Q_x
& u\tau_{yx}+v\tau_{yy}+w\tau_{yz}-Q_y
& u\tau_{zx}+v\tau_{zy}+w\tau_{zz}-Q_z\\
-J_{x,1} & -J_{y,1} & -J_{z,1}\\
\vdots & \vdots & \vdots\\
-J_{x,N-1} & -J_{y,N-1} & -J_{z,N-1}
\end{array}
\right]
$}
\end{gathered}
\end{gathered}
\label{eq:eq2}
\end{equation}

\noindent where $\rho$ and $p$ denote the density and pressure of the gas mixture, respectively.
$u$, $v$, and $w$ are the velocity components. $\sigma$, $\kappa$, $\delta_\Gamma$, and
$\bm{n}$ denote the surface-tension coefficient, interface curvature, Dirac delta function
supported on $\Gamma(t)$, and unit interface-normal vector, respectively. $Y_k$ represents the
mass fraction of species $k$ ($k=1,\cdots,N-1$), and the mass fraction of the $N$-th species is obtained from

\begin{equation}
Y_N=1-\sum_{k=1}^{N-1}Y_k.
\label{eq:eq4}
\end{equation}
$E$ is the specific total energy, related to the primitive variables by

\begin{equation}
E=h-\frac{p}{\rho}+\frac{u^2+v^2+w^2}{2},
\label{eq:eq5}
\end{equation}

\noindent where $h$ is the specific enthalpy of the gas mixture.

The viscous stress tensor in Eq.~\eqref{eq:eq2} is modeled for a Newtonian fluid under Stokes'
hypothesis, with $\mu$ taken as the mixture-averaged viscosity. The mass diffusion flux of
species $k$ in Eq.~\eqref{eq:eq2} is

\begin{equation}
J_{i,k}=\rho Y_k V_{i,k}^d,
\label{eq:eq7}
\end{equation}

\noindent where $V_{i,k}^d$ denotes the diffusion velocity of species $k$ in the $i$-th
direction. Neglecting the Soret effect \cite{houim2013ghostFluidPhaseChange}, the diffusion
velocities are first calculated by

\begin{equation}
\tilde{V}_{i,k}^d
=-\frac{D_{k,\mathrm{mix}}}{X_k}
\left(\frac{\partial X_k}{\partial x_i}
+(X_k-Y_k)\frac{\partial \ln p}{\partial x_i}\right),
\label{eq:eq8}
\end{equation}

\noindent where $X_k$ is the mole fraction and $x_i$ denotes the $i$-th Cartesian coordinate.
The mixture-averaged diffusion coefficient $D_{k,\mathrm{mix}}$ of species $k$ is obtained from

\begin{equation}
D_{k,\mathrm{mix}}
=\frac{1-Y_k}{\displaystyle\sum_{l=1,l\ne k}^{N}\frac{X_l}{D_{k,l}}},
\label{eq:eq9}
\end{equation}

\noindent where $D_{k,l}$ is the binary diffusion coefficient. To ensure mass conservation, the
diffusion velocity is corrected as

\begin{equation}
V_{i,k}^d=\tilde{V}_{i,k}^d-\sum_{l=1}^{N}Y_l\tilde{V}_{i,l}^d.
\label{eq:eq10}
\end{equation}

\noindent The diffusive energy flux contribution $Q_i$ in Eq.~\eqref{eq:eq2} is calculated by

\begin{equation}
Q_i=\sum_{k=1}^{N}J_{i,k}h_k-\lambda \frac{\partial T}{\partial x_i},
\label{eq:eq11}
\end{equation}

\noindent where the first and second terms on the right-hand side represent the effects of
species and thermal diffusion, respectively. $h_k$ is the specific enthalpy of species $k$ and
$\lambda$ is the mixture-averaged thermal conductivity. The mixing rules of transport
coefficients of both viscosity $\mu$ and thermal conductivity $\lambda$ follow Wilke's rule
\cite{wilke1950viscosityMixtures}.

\subsection{Equations of state}
\label{sec:eos}

To close the governing equations, the thermodynamic relationship between the pressure, density,
and internal energy for each phase must be defined. The gas phase is modeled as a mixture of $N$
species of ideal gases. The pressure $p$ of the gas phase is determined by the summation of
partial pressures $p_k$ according to Dalton's law

\begin{equation}
p=\sum_{k=1}^{N}p_k=\rho R_uT\sum_{k=1}^{N}\frac{Y_k}{M_{w,k}},
\label{eq:eq14}
\end{equation}

\noindent where $R_u$ is the universal gas constant and $M_{w,k}$ is the molar mass of
species $k$. Assuming that the thermally perfect gas mixture is in local thermodynamic
equilibrium, the temperature of the gas phase $T$ is determined by solving the specific-energy
relation for $T$ using the Newton-Raphson method

\begin{equation}
e(T)=\sum_{k=1}^{N}\left(Y_kh_k(T)-\frac{R_u}{M_{w,k}}T\right).
\label{eq:eq15}
\end{equation}

For the liquid materials considered in the present test cases, water and aluminum are described
by the stiffened-gas EOS \cite{lemetayer2004eosLiquidVapor} and the Tait EOS
\cite{das2020shockVaporizationDroplets}, respectively. For
the stiffened-gas EOS, the relations between different thermodynamic variables are

\begin{equation}
\begin{aligned}
e(p,\rho)&=\frac{p+\gamma p_{\infty}}{(\gamma-1)\rho}+e_{\mathrm{ref}},\\
T(p,\rho)&=\frac{p+p_{\infty}}{C_v(\gamma-1)\rho},
\end{aligned}
\label{eq:eq16}
\end{equation}

\noindent where $\gamma$ is the adiabatic coefficient, $p_{\infty}$ is the parameter accounting
for the pre-compression of the fluid, $e_{\mathrm{ref}}$ is the reference internal energy, and
$C_v$ is the heat capacity at constant volume. Following
Ref.~\cite{lemetayer2004eosLiquidVapor}, we choose $\gamma=2.35$, $p_{\infty}=10^9~\mathrm{Pa}$,
$C_v=1.816\times10^3~\mathrm{J/(kg\cdot K)}$ for liquid water. The reference internal energy is
set to $e_{\mathrm{ref}}=-1.713\times10^7~\mathrm{J/kg}$, calibrated against the JANAF
thermodynamic data \cite{browne2018sdtoolbox} for water vapor.

For liquid aluminum, the Tait EOS is adopted to describe the relationship between pressure and
density

\begin{equation}
p(\rho)=B\left[\left(\frac{\rho}{\rho_0}\right)^\gamma-1\right]+p_0,
\label{eq:eq17}
\end{equation}

\noindent where $\gamma$ is an artificial specific-heat ratio, $\rho_0$ and $p_0$ are reference
density and pressure, respectively. $B$ is a material-dependent constant related to the bulk
modulus. Based on the study by Das and Udaykumar~\cite{das2020shockVaporizationDroplets}, the
parameters for liquid aluminum are set to $\gamma=8.55$, $\rho_0=2003~\mathrm{kg/m^3}$, and
$p_0=10^5~\mathrm{Pa}$, with $B=3.36\times10^9~\mathrm{Pa}$ in this work. The Tait EOS provides a
mechanical closure for liquid aluminum, and the temperature is obtained from a caloric relation
with a constant specific heat capacity $C_v$
\cite{houim2013ghostFluidPhaseChange}

\begin{equation}
e(T)=e_{\mathrm{ref}}+C_v(T-T_{\mathrm{ref}}),
\label{eq:eq18}
\end{equation}

\noindent where $e_{\mathrm{ref}}$ and $T_{\mathrm{ref}}$ are the reference specific energy and
temperature of the liquid, respectively. For liquid aluminum, we use
$C_v=1177~\mathrm{J/(kg\cdot K)}$, and the reference value is set to
$e_{\mathrm{ref}}=-2.926\times10^6~\mathrm{J/kg}$ at
$T_{\mathrm{ref}}=2750~\mathrm{K}$.

The liquid reference energies are calibrated against the gas-phase thermochemical enthalpy of
the phase-changing species at fixed material reference states, with
$T_{\mathrm{cal}}=373.15~\mathrm{K}$ for water and $2750~\mathrm{K}$ for aluminum. The
corresponding reference pressures are the saturation pressures at these temperatures. The
calibration enforces

\[
h_{\mathrm{vap}}(T_{\mathrm{cal}},p_{\mathrm{sat}}(T_{\mathrm{cal}}))
-h_{\mathrm{liq}}(T_{\mathrm{cal}},p_{\mathrm{sat}}(T_{\mathrm{cal}}))
=Q_{\mathrm{lat}},
\]

\noindent where $h_{\mathrm{vap}}$ is evaluated from the gas-phase thermochemical data and
$h_{\mathrm{liq}}$ from the liquid EOS. Thus, the liquid and gas thermodynamic descriptions use
a consistent energy reference at the calibration state.

\subsection{Chemical reaction model}

Gas-phase chemistry is modeled using a finite-rate chemical kinetics formulation.
The chemical evolution in the reactive sub-step is described by a generic multi-step mechanism
involving $N$ species and $N_R$ elementary reactions

\begin{equation}
\sum_{k=1}^{N}\nu'_{k,j}\mathcal{X}_k
\underset{k_{b,j}}{\stackrel{k_{f,j}}{\rightleftharpoons}}
\sum_{k=1}^{N}\nu''_{k,j}\mathcal{X}_k
\quad (j=1,2,\ldots,N_R),
\label{eq:eq19}
\end{equation}

\noindent where $\mathcal{X}_k$ represents the chemical symbol of species $k$, and $\nu'_{k,j}$ and
$\nu''_{k,j}$ are the stoichiometric coefficients for the reactants and products, respectively.
The net mass production rate for species $k$ is obtained by summing the contributions from all
reactions

\begin{equation}
\dot{\omega}_k=M_{w,k}\sum_{j=1}^{N_R}(\nu''_{k,j}-\nu'_{k,j})
\left[
k_{f,j}\prod_{s=1}^{N}C_s^{\nu'_{s,j}}
-k_{b,j}\prod_{s=1}^{N}C_s^{\nu''_{s,j}}
\right],
\label{eq:eq20}
\end{equation}

\noindent where $C_s$ is the molar concentration of species $s$. The forward rate coefficient $k_{f,j}$ is
determined by the modified Arrhenius law \cite{houim2013ghostFluidPhaseChange}

\begin{equation}
k_{f,j}=A_jT^{b_j}\exp\left(-\frac{E_{a,j}}{R_uT}\right),
\label{eq:eq21}
\end{equation}

\noindent while the reverse rate coefficient is calculated from the equilibrium constant,
$k_{b,j}=k_{f,j}/K_{c,j}$. The concentration-based equilibrium constant $K_{c,j}$ is evaluated
from the standard-state thermodynamic data of the participating species,

\begin{equation}
K_{c,j}=\left(\frac{p_{\mathrm{atm}}}{R_uT}\right)^{\Delta\nu_j}
\exp\left(-\frac{\Delta G_j^0(T)}{R_uT}\right),
\qquad
\Delta\nu_j=\sum_{k=1}^{N}\left(\nu''_{k,j}-\nu'_{k,j}\right),
\label{eq:eq22}
\end{equation}

\noindent where $p_{\mathrm{atm}}$ is the standard-state pressure, and $\Delta G_j^0(T)$ is the
standard molar Gibbs free-energy change of reaction $j$, computed from the NASA polynomial
thermodynamic data \cite{mcbride2002nasaGlennCoefficients}.

\subsection{Conservative discretization}

\begin{figure}[!htbp]
  \centering
  \includegraphics[width=0.6\linewidth]{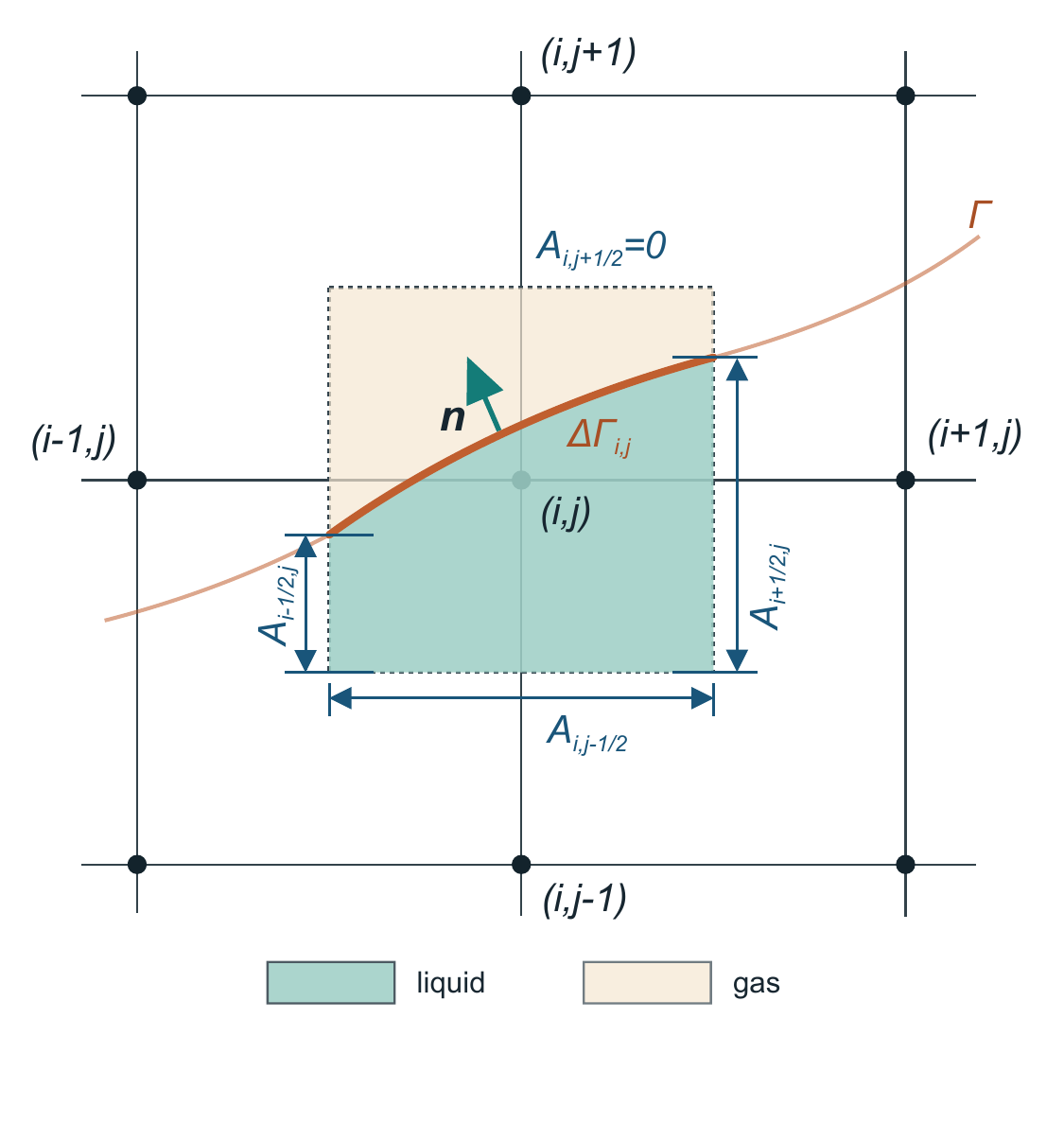}
  \caption{2D schematic of the conservative discretization in a cut cell $(i,j)$.
    The colored regions denote the portions of the cut cell occupied by the liquid and gas
    phases. The thick interface segment $\Delta\Gamma_{i,j}$ is used for the local interfacial
    flux evaluation, while the lighter extension indicates the interface continuation through
    neighboring cells. The arrow denotes the unit normal vector $\bm{n}$ at the interface.
    Symbols $A_{i\pm1/2,j}$ and $A_{i,j\pm1/2}$ denote the corresponding face apertures, i.e.,
    the effective face lengths available for flux exchange across each side of the cut cell.}
  \label{fig:conservative-discretization}
\end{figure}

In this work, the governing equations for compressible two-phase, multi-species flows are solved
using a finite-volume method on a Cartesian grid. To maintain the conservative
property, a conservative sharp-interface method \cite{hu2006conservativeInterface} is employed,
whose key idea is to modify the spatial discretization in cells cut by the interface. The
two-dimensional (2D) schematic of the conservative discretization in a cut cell
$(i,j)$ is shown in Fig.~\ref{fig:conservative-discretization}. The second-order Strang splitting
scheme \cite{strang1968operatorSplitting} is used to separate
chemical-source integration from fluid transport. The chemical source terms are integrated using
the CHEMEQ2 algorithm \cite{mott2001chemeq2}. During the transport sub-step, Eq.~\eqref{eq:eq1}
is discretized by the first-order forward Euler method as follows

\begin{equation}
\begin{aligned}
\alpha_{i,j}^{n+1}\bm{U}_{i,j}^{n+1}
={}&\alpha_{i,j}^{n}\bm{U}_{i,j}^{n}
+\frac{\Delta t}{\Delta x\Delta y}
\left[\hat{\bm{X}}(\Delta\Gamma_{i,j})
+\hat{\bm{X}}_v(\Delta\Gamma_{i,j})\right]
\\
&+\frac{\Delta t}{\Delta x}
\left[A_{i-1/2,j}\hat{\bm{F}}_{i-1/2,j}
-A_{i+1/2,j}\hat{\bm{F}}_{i+1/2,j}\right]
\\
&+\frac{\Delta t}{\Delta y}
\left[A_{i,j-1/2}\hat{\bm{F}}_{i,j-1/2}
-A_{i,j+1/2}\hat{\bm{F}}_{i,j+1/2}\right]
\\
&+\frac{\Delta t}{\Delta x}
\left[A_{i-1/2,j}\hat{\bm{F}}_{v,i-1/2,j}
-A_{i+1/2,j}\hat{\bm{F}}_{v,i+1/2,j}\right]
\\
&+\frac{\Delta t}{\Delta y}
\left[A_{i,j-1/2}\hat{\bm{F}}_{v,i,j-1/2}
-A_{i,j+1/2}\hat{\bm{F}}_{v,i,j+1/2}\right],
\end{aligned}
\label{eq:eq23}
\end{equation}

\noindent where $\Delta t$ is the time step and $\alpha_{i,j}\in[0,1]$ is the volume fraction of
cell $(i,j)$ occupied by the considered phase, so that $\alpha_{i,j}\bm{U}_{i,j}$ are the
phase-conservative quantities stored in the cut cell. $\hat{\bm{X}}(\Delta\Gamma_{i,j})$ and
$\hat{\bm{X}}_v(\Delta\Gamma_{i,j})$ denote the inviscid and viscous fluxes across the interface,
respectively. The signs of the viscous contributions are included in the definitions of the
numerical fluxes, following the conservative-interface convention
\cite{luo2015conservativeSharpInterface,long2021acceleratedConservativeSharpInterface}.

To evaluate the viscous flux $\hat{\bm{X}}_v(\Delta\Gamma_{i,j})$ across the interface, an
effective viscosity approach is implemented within the cut cells based on the assumptions of a
locally linear velocity distribution within each fluid and continuous viscous shear stress
across the phase interface \cite{luo2015conservativeSharpInterface}. The inviscid flux
$\hat{\bm{X}}(\Delta\Gamma_{i,j})$ is obtained by solving a two-phase multi-species Riemann
problem with phase change and surface tension along the normal direction $\bm{n}$, which is the
focus of this work. The thermal effect associated with mass transfer is represented by the latent
heat term in the interfacial Riemann problem. In the present model, the Fourier heat flux driven
by a temperature gradient across the phase interface is assumed negligible relative to the
latent-heat exchange associated with phase change and is therefore omitted. Heat conduction is
evaluated separately within each phase from its bulk temperature gradient.

\subsection{Interface coupling method}

The liquid--gas interface $\Gamma(t)$ is resolved sharply to define the cut-cell geometry. The
gas-side mixture state, obtained from the diffuse-interface treatment of species transport and
chemical reactions, is evaluated adjacent to $\Gamma(t)$. Together with the adjacent liquid state,
it defines a local phase-change Riemann problem, whose solution yields the interfacial flux and
interface velocity.
For convenience, we define the interface-normal state vectors used in the Riemann problem for the
gas and liquid phases separately

\begin{equation}
\bm{W}_g=(\rho_g,v_g,p_g,e_g,Y_1,\ldots,Y_{N-1})^T\in\mathbb{R}^{3+N},
\quad
\bm{W}_l=(\rho_l,v_l,p_l,e_l)^T\in\mathbb{R}^4,
\label{eq:eq24}
\end{equation}

\noindent where $N$ is the total number of species in the gas phase, and $v=\bm{u}\cdot\bm{n}$
denotes the normal velocity. The interface-normal Riemann problem is then written as

\begin{equation}
\mathcal{R}_{i,j}=
\begin{cases}
\mathcal{R}(\bm{W}_{i,j}^{l},\bm{W}_{i,j}^{g,G}), & \text{if }(i,j)\in\Omega_l,\\
\mathcal{R}(\bm{W}_{i,j}^{l,G},\bm{W}_{i,j}^{g}), & \text{if }(i,j)\in\Omega_g,
\end{cases}
\label{eq:eq25}
\end{equation}

\noindent where the superscript $G$ indicates the ghost states obtained by using the extension
algorithm \cite{fu2017levelSetReinitialization}.

The interface is captured using the level-set method \cite{osher1988levelSet}, where the sharp
interface between the liquid and the gas phases is implicitly represented by a signed distance
function $\phi(\bm{x},t)$. The time evolution of $\phi(\bm{x},t)$ is governed by

\begin{equation}
\frac{\partial \phi}{\partial t}+\hat{\bm{u}}\cdot\nabla\phi=0,
\label{eq:eq26}
\end{equation}

\noindent where $\hat{\bm{u}}$ represents the interface velocity obtained from the interfacial
Riemann solver. The unit normal vector $\bm{n}$ and the local curvature $\kappa$ are directly
derived from

\begin{equation}
\bm{n}=\frac{\nabla\phi}{|\nabla\phi|},\qquad
\kappa=\nabla\cdot\frac{\nabla\phi}{|\nabla\phi|},
\label{eq:eq27}
\end{equation}

\noindent where the signed-distance property $|\nabla\phi|=1$ is maintained by the
reinitialization procedure of Ref.~\cite{fu2017levelSetReinitialization}. In this work,
$\phi>0$ denotes the gas phase and $\phi<0$ denotes the liquid phase, so that $\bm{n}$ points
from the liquid to the gas. The unit normal $\bm{n}$ defines the direction of the
interface-normal Riemann problem, and the curvature $\kappa$ enters the surface-tension
contribution to the interfacial jump conditions introduced in Section~\ref{sec:riemann-problem}.

\section{Two-phase multi-species Riemann problem with phase change}
\label{sec:riemann-problem}

\subsection{Solution structure}

As described above, the interfacial flux in a cut cell is obtained by constructing and solving a
one-dimensional (1D) two-phase Riemann problem in the interface-normal direction. In this work, we
construct an approximate multi-species Riemann solver by extending the original single-species
four-wave phase-change Riemann solver \cite{long2023conservativePhaseChange} to a gas mixture.
For the multi-species system, the additional species equations introduce convective
characteristic fields associated with the flow velocity, which are linearly degenerate
\cite{larrouturou1989multicomponentGas}. Thus, species mass fractions may jump across the
contact wave, but they do not introduce additional acoustic or phase-interface waves. As a
result, the multi-species
phase-change Riemann problem retains the four-wave structure of the single-species Riemann problem
\cite{long2023conservativePhaseChange}, while the interfacial closure relations must be
modified to account for species-selective mass and energy transfer.

Without loss of generality, the liquid phase is placed on the left side of the interface and the
gas mixture on the right side. The initial states of the local Riemann problem are written as

\begin{equation}
\bm{W}_L=(\rho_L,v_L,p_L,e_L)^T,\qquad
\bm{W}_R=(\rho_R,v_R,p_R,e_R,Y_1,\ldots,Y_{N-1})^T.
\label{eq:eq28}
\end{equation}

\begin{figure}[!htbp]
  \centering
  \includegraphics[width=\linewidth]{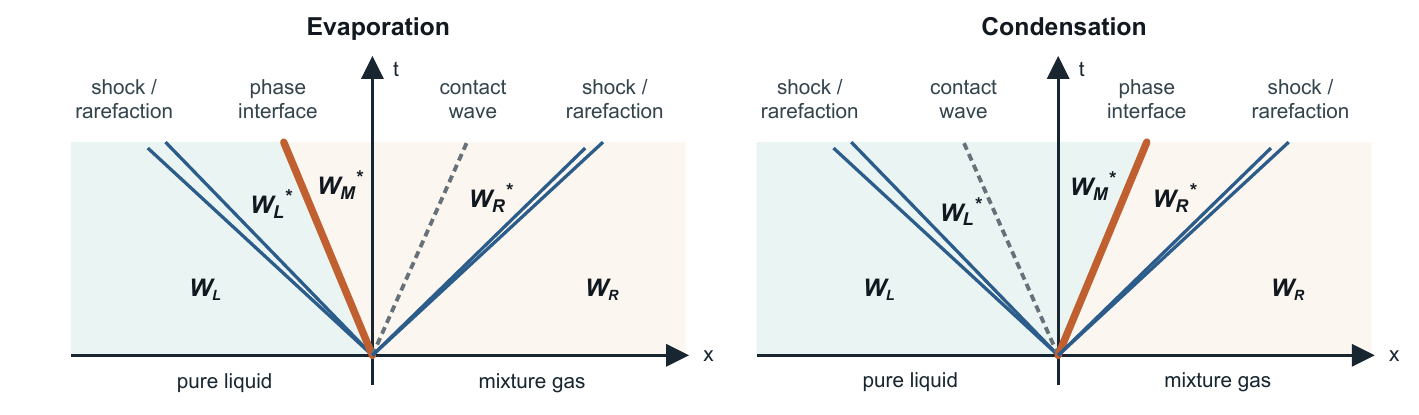}
  \caption{The wave structures of the multi-species Riemann problem with evaporation (left) and
    condensation (right).}
  \label{fig:riemann-wave-structure}
\end{figure}

\noindent Here, $\bm{W}_L$ and $\bm{W}_R$ correspond to the liquid phase and the gas mixture,
respectively. $Y_1$ denotes the condensable vapor species associated with the liquid phase, and
it is the only gas species that can exchange mass with the liquid across the phase interface. As
shown in Fig.~\ref{fig:riemann-wave-structure}, the four-wave Riemann fan divides the solution into three
intermediate states, denoted by $\bm{W}_L^*$, $\bm{W}_M^*$, and $\bm{W}_R^*$. The outer acoustic
waves connect $\bm{W}_L$ to $\bm{W}_L^*$ and $\bm{W}_R^*$ to $\bm{W}_R$, respectively. The inner
waves consist of the phase interface and the contact wave, whose relative ordering depends on
the direction of phase change. The corresponding Riemann wave patterns for evaporation and
condensation are illustrated in Fig.~\ref{fig:riemann-wave-structure}. The jump conditions and closure relations
used to determine these intermediate states are presented in the next subsection.

\subsection{Approximate Riemann solver for the multi-species Riemann problem with phase change}
\label{sec:riemann-solver}

\begin{figure}[!htbp]
  \centering
  \includegraphics[width=\linewidth]{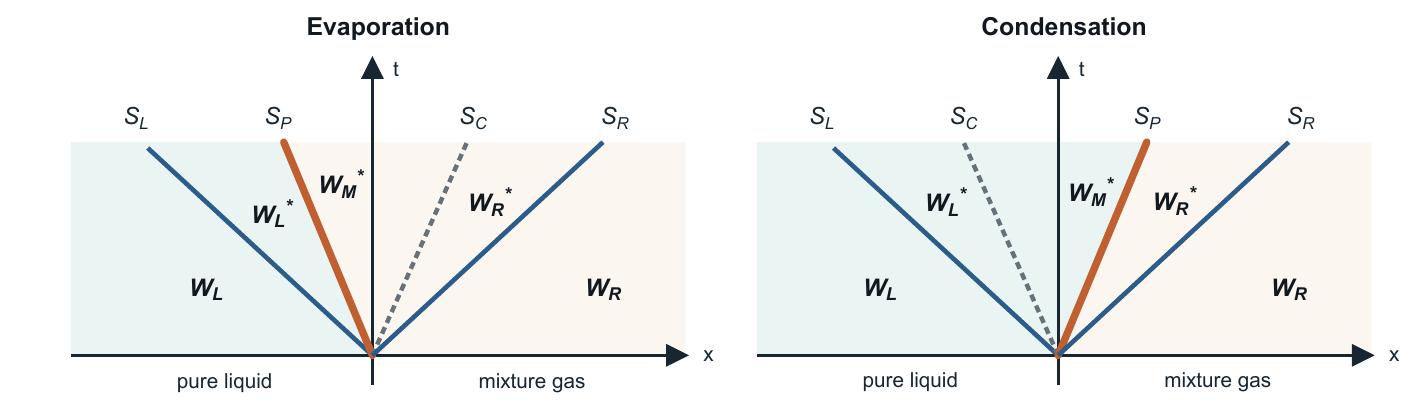}
  \caption{The approximate solution structure of the multi-species Riemann problem with
    evaporation (left) and condensation (right).}
  \label{fig:hllc-riemann-structure}
\end{figure}

Based on the solution structure described above, we develop a new approximate Riemann solver for
the multi-species phase-change problem. The solver follows the four-wave structure of the
single-species solver \cite{long2023conservativePhaseChange}, while the closure relations
and the phase-interface jump conditions are modified to provide a species-selective interfacial
energy balance for a multi-species gas phase. As shown in Fig.~\ref{fig:hllc-riemann-structure}, the HLLC-type approximate solution
consists of five constant states separated by four waves: two outer acoustic waves, a phase
interface, and a contact wave. In the spirit of HLLC-type solvers, the outer waves, which may be
shocks or rarefactions in the exact solution, are approximated as single discontinuities, and
their speeds are estimated following Ref.~\cite{long2023conservativePhaseChange},

\begin{equation}
S_L=v_L-c_L,\qquad S_R=v_R+c_R.
\label{eq:eq29}
\end{equation}

Across the outer waves, the standard Rankine-Hugoniot jump conditions
\cite{toro2009riemannSolvers} are enforced

\begin{equation}
\begin{aligned}
{[\rho(v-S)]}&=0,\\
{[\rho(v-S)v+p]}&=0,\\
\left[\rho(v-S)\left(e+\frac{1}{2}v^2\right)+pv\right]&=0,
\end{aligned}
\label{eq:eq30}
\end{equation}

\noindent where $S$ is the wave speed and $[a]=a_R-a_L$ denotes the jump operator, defined as the
right state minus the left state. On the gas side, the species equations additionally give

\begin{equation}
[\rho Y_k(v-S)]=0,\qquad k=1,\ldots,N-1.
\label{eq:eq31}
\end{equation}

\noindent For the contact wave, the jump conditions reduce to

\begin{equation}
v_M^*=S_C,\qquad [p]=0,\qquad [v]=0,
\label{eq:eq32}
\end{equation}

\noindent while the density, specific internal energy, and species mass fractions may be discontinuous
across the contact wave.

Next, we consider the jump conditions across the phase interface. Let $S_P$ denote the velocity
of the phase interface. The mass flux for phase change, $j$, denotes the net mass exchange
between the liquid and gas phases. The total-mass balance gives

\begin{equation}
j=\rho_{\mathrm{liq}}^*(v_{\mathrm{liq}}^*-S_P)
=\rho_{\mathrm{gas}}^*(v_{\mathrm{gas}}^*-S_P),
\label{eq:eq33}
\end{equation}

\noindent The subscripts $\mathrm{liq}$ and $\mathrm{gas}$ denote the liquid and gas-mixture states
adjacent to the phase interface, respectively. For reference, the net normal flux of gas species
$k$ relative to the moving interface has the standard convective--diffusive decomposition

\begin{equation}
j_k=\rho_{\mathrm{gas}}^*Y_k^*(v_{\mathrm{gas}}^*-S_P)+J_{k,n}^*,
\qquad k=1,\ldots,N,
\label{eq:eq34}
\end{equation}

\noindent where $J_{k,n}^*$ is the normal diffusive contribution. The first term represents
convective transport relative to the moving interface, and the mixture-averaged diffusion model
satisfies $\sum_{k=1}^{N}J_{k,n}^*=0$. Summing Eq.~\eqref{eq:eq34} over all species therefore
recovers Eq.~\eqref{eq:eq33}. With the normal direction from the liquid to the gas, positive
$j$ corresponds to mass transfer from the liquid to the gas. Their correspondence with the
intermediate states depends on the sign of $j$

\begin{equation}
(\mathrm{liq},\mathrm{gas})=
\begin{cases}
(L,M), & j>0\ \text{(evaporation)},\\
(M,R), & j<0\ \text{(condensation)}.
\end{cases}
\label{eq:eq35}
\end{equation}

The phase interface is assumed to be subsonic and non-characteristic, following the standard
treatment of phase-change Riemann problems \cite{long2023conservativePhaseChange}. Using the
liquid-side and gas-side states defined above, the jump conditions of mass and momentum balance
across the phase interface are written as

\begin{equation}
\begin{aligned}
{[j]}&=0,\\
{[jv+p]}&=-\sigma\kappa.
\end{aligned}
\label{eq:eq36}
\end{equation}

\noindent Since only the condensable vapor species crosses the phase interface, the present
conservative coupling directly prescribes the net species exchanges as

\begin{equation}
 j_1=j,\qquad j_k=0,\qquad k=2,\ldots,N.
\label{eq:eq37}
\end{equation}

\noindent Thus, only the condensable vapor has a nonzero net interfacial flux, and
$\sum_{k=1}^{N}j_k=j$. The cut-cell discretization directly imposes these net exchange fluxes,
which already account for interfacial species transfer, and therefore does not require an
additional gas-side species diffusion-flux boundary condition.

For the energy balance, a natural extension of the
single-species formulation is to use the gas-mixture internal energy in the interfacial energy
jump condition. However, this treatment is inconsistent with species-selective phase change,
because only the condensable vapor crosses the phase interface. 
Let $e_k$ denote the specific internal energy of species $k$, so that the gas-mixture
internal energy is $e=\sum_{k=1}^{N}Y_k e_k$. Under the single-velocity assumption, all gas
species share the same velocity $v$. Hence, the energy flux associated with phase change is

\begin{equation}
\mathcal{F}_E^{pc}
=\sum_{k=1}^{N}j_k\left(e_k+\frac{1}{2}v^2\right)
=j\left(e_1+\frac{1}{2}v^2\right).
\label{eq:eq38}
\end{equation}

\noindent The mixture-based treatment would instead give

\begin{equation}
j\left(e+\frac{1}{2}v^2\right)
=\sum_{k=1}^{N}jY_k\left(e_k+\frac{1}{2}v^2\right),
\label{eq:eq39}
\end{equation}

\noindent which is equivalent to replacing the physically admissible interfacial species-flux vector

\begin{equation}
(j_1,j_2,\ldots,j_N)=(j,0,\ldots,0),
\label{eq:eq40}
\end{equation}

\noindent with the fictitious mixture-flux vector

\begin{equation}
(\tilde{j}_1,\tilde{j}_2,\ldots,\tilde{j}_N)
=(jY_1,jY_2,\ldots,jY_N).
\label{eq:eq41}
\end{equation}

\noindent This fictitious flux would transport non-condensable species across the phase interface despite
their actual phase-change fluxes being zero. Consequently, using the gas-mixture internal energy
in the interfacial energy jump condition introduces an unphysical energy exchange. As will be
shown in Sec.~\ref{sec:impulsive-condensation}, this inconsistency leads to pronounced spurious
interfacial artifacts, including sharp pressure spikes and an unphysical temperature depression
near the phase interface. Thus, we define

\begin{equation}
e_{pc}=
\begin{cases}
e_{\mathrm{liq}}, & \text{on the liquid side},\\
e_1, & \text{on the gas side},
\end{cases}
\label{eq:eq42}
\end{equation}

\noindent where $e_1$ is the specific internal energy of the condensable vapor species in the
gas-side state adjacent to the phase interface. To account for the species-selective interfacial
energy transfer, the energy jump condition is rewritten as

\begin{equation}
\left[j\left(e_{pc}+\frac{1}{2}v^2\right)+pv\right]
=jQ_{\mathrm{lat}}-\sigma\kappa S_P,
\label{eq:eq43}
\end{equation}

\noindent where $Q_{\mathrm{lat}}$ is the specific latent heat of vaporization of the
phase-changing species, treated as a constant in this work and set to
$2.242\times10^{6}~\mathrm{J/kg}$ for water and
$1.704\times10^{7}~\mathrm{J/kg}$ for aluminum.
The term $jQ_{\mathrm{lat}}$ appears only in the interfacial flux balance. It represents
the phase-change energy associated with material transfer across the interface. The values of
$Q_{\mathrm{lat}}$ are consistent with the liquid reference-energy calibration in
Sec.~\ref{sec:eos}. They do not introduce an additional phase-energy offset already
contained in the equations of state.
Together with the species-selective definition of $e_{pc}$ in Eq.~\eqref{eq:eq42},
Eq.~\eqref{eq:eq43} provides the multi-species energy jump used in the interfacial Riemann
problem.

With the jump conditions of Eqs.~\eqref{eq:eq30}, \eqref{eq:eq31}, \eqref{eq:eq32},
\eqref{eq:eq36}, \eqref{eq:eq43} and the total-mass flux relation Eq.~\eqref{eq:eq33}, an algebraic
system for the approximate Riemann solution is obtained. The number of unknowns and equations
depend on the direction of phase change. For evaporation, the unknowns are

\begin{equation}
\begin{gathered}
\bm{W}_L^*=(\rho_L^*,v_L^*,p_L^*,e_L^*)^T,\\
\bm{W}_M^*=(\rho_M^*,v_M^*,p_M^*,e_M^*,Y_{1,M}^*,\ldots,Y_{N-1,M}^*)^T,\\
\bm{W}_R^*=(\rho_R^*,v_R^*,p_R^*,e_R^*,Y_{1,R}^*,\ldots,Y_{N-1,R}^*)^T,
\end{gathered}
\label{eq:eq44}
\end{equation}

\noindent together with $S_P$, $S_C$, and $j$, giving $15+2(N-1)$ unknowns. The hydrodynamic jump
conditions across the four waves, together with the definition of $j$, provide 13 independent
equations. The species jump conditions across the gas-side outer wave provide $N-1$ additional
equations for $Y_{k,R}^*$. Thus, $13+(N-1)$ equations are available before the closure relations
are introduced. In evaporation, the contact wave does not determine the species mass fractions
in $\bm{W}_M^*$. We therefore close the approximate Riemann problem by taking the species mass
fractions of this gas-side intermediate state from the adjacent gas state:

\begin{equation}
Y_{k,M}^*=Y_{k,R}^*,\qquad k=1,\ldots,N-1.
\label{eq:eq45}
\end{equation}

\noindent Both $\bm{W}_M^*$ and $\bm{W}_R^*$ are gas-phase intermediate states separated by the
contact wave. Equation~\eqref{eq:eq45} is an approximate closure that assumes no species
mass-fraction jump across this gas-phase contact. It supplies the remaining $N-1$ species
mass-fraction relations, so two additional
closure relations are still required. For condensation, $\bm{W}_M^*$ corresponds to a
liquid-phase state and contains no gas species mass fractions. Therefore, compared with
evaporation, the $N-1$ composition unknowns in $\bm{W}_M^*$ are removed, and the total number of
unknowns becomes $15+(N-1)$. The gas-side species jump conditions determine $Y_{k,R}^*$, again
leaving two closure relations to be specified.

The density-ratio closure of Long et al.~\cite{long2023conservativePhaseChange} is adopted by
assuming that the density ratio across the phase interface remains approximately invariant
between the initial and intermediate states. The closure is written in a unified form as

\begin{equation}
\frac{\rho_{\mathrm{liq}}^*}{\rho_{\mathrm{gas}}^*}
\approx \frac{\rho_L}{\rho_R},
\label{eq:eq46}
\end{equation}

\noindent where $\rho_{\mathrm{liq}}^*$ and $\rho_{\mathrm{gas}}^*$ denote the liquid-side and
gas-side intermediate densities adjacent to the phase interface, respectively. Specifically,
$\rho_{\mathrm{liq}}^*=\rho_L^*$ and $\rho_{\mathrm{gas}}^*=\rho_M^*$ for evaporation, while
$\rho_{\mathrm{liq}}^*=\rho_M^*$ and $\rho_{\mathrm{gas}}^*=\rho_R^*$ for condensation.
This is the direct multi-species extension of the closure in Long et
al.~\cite{long2023conservativePhaseChange}. In the gas mixture,
$\rho_{\mathrm{gas}}^*=\sum_{k=1}^{N}\rho_k^*$ with
$\rho_k^*=\rho_{\mathrm{gas}}^*Y_k^*$, and the total-mass jump condition in
Eq.~\eqref{eq:eq33} involves $\rho_{\mathrm{gas}}^*$ rather than the partial density of the
condensable species. Freezing the liquid-to-gas density ratio therefore gives
Eq.~\eqref{eq:eq46}. In the single-species limit, $Y_1=1$ and
$\rho_{\mathrm{gas}}^*=\rho_{\mathrm{vap}}^*$, so that it recovers the original pure-fluid
closure exactly.

The second closure relation is supplied by a mass flux model for phase change. Here, we use the
Schrage--Knudsen equation of Schrage~\cite{schrage1953interphaseMassTransfer}, in the form
adopted by Houim and Kuo~\cite{houim2013ghostFluidPhaseChange}. The mass flux is evaluated
with the intermediate states adjacent to the phase interface for numerical consistency as
proposed by Long et al.~\cite{long2023conservativePhaseChange}:

\begin{equation}
j=\frac{2a}{2-a}\sqrt{\frac{M_{w,1}}{2\pi R_u}}
\left(
\frac{p_{\mathrm{sat}}(T_{\mathrm{liq}}^*)}{\sqrt{T_{\mathrm{liq}}^*}}
-\frac{p_{\mathrm{vap}}^*}{\sqrt{T_{\mathrm{gas}}^*}}
\right),
\label{eq:eq47}
\end{equation}

\noindent where $p_{\mathrm{vap}}^*=p_{\mathrm{gas}}^*X_1^*$ is the condensable-vapor partial
pressure in the gas-side state adjacent to the phase interface.
Following Houim and Kuo~\cite{houim2013ghostFluidPhaseChange}, the phase-change driving force
is determined by the condensable-vapor partial pressure relative to the saturation pressure,
rather than by the direct contribution of non-condensable species to the total gas pressure.
$M_{w,1}$ is the molar mass of the condensable vapor species.
$R_u$ is the universal gas constant. The accommodation coefficient $a$ is evaluated using the
model of Nagayama and Tsuruta~\cite{nagayama2003condensationCoefficient} with the
condensable-vapor partial density $\rho_{\mathrm{vap}}^*=\rho_{\mathrm{gas}}^*Y_1^*$ as

\begin{equation}
a=\left\{1-\left(\frac{\rho_{\mathrm{vap}}^*}{\rho_{\mathrm{liq}}^*}\right)^{1/3}\right\}
\exp\left[
-\frac{1}{2\left(\frac{\rho_{\mathrm{liq}}^*}{\rho_{\mathrm{vap}}^*}\right)^{1/3}-2}
\right].
\label{eq:eq48}
\end{equation}

\noindent The saturation pressure $p_{\mathrm{sat}}(T_l)$ can be evaluated from material-dependent
correlations. For water, the correlation used in Ref.~\cite{long2023conservativePhaseChange} is
adopted:

\begin{equation}
\begin{aligned}
p_{\mathrm{sat}}(T_l)=611.2\exp(&1045.8511577-21394.6662629T_l^{-1}
+1.0969044T_l\\
&-1.3003741\times10^{-3}T_l^2
+7.7472984\times10^{-7}T_l^3\\
&-2.1649005\times10^{-12}T_l^4
-211.3896559\ln T_l).
\end{aligned}
\label{eq:eq49}
\end{equation}

\noindent For aluminum, we use the equation in Ref.~\cite{das2021reactingAluminumDroplets}:

\begin{equation}
p_{\mathrm{sat}}(T_l)=\exp\left(36.547-\frac{39033}{T_l}
-1.3981\ln T_l+6.7839\times10^{-9}T_l^2\right).
\label{eq:eq50}
\end{equation}

\noindent In Eqs.~\eqref{eq:eq49} and \eqref{eq:eq50}, $T_l$ is given in Kelvin and
$p_{\mathrm{sat}}$ is obtained in Pa.

To simplify the nonlinear iterative system and improve convergence robustness, we adopt the
reduction strategy of Long et al.~\cite{long2023conservativePhaseChange}, which parametrizes
the algebraic system by the mass flux for phase change, $j$. All
intermediate states and wave speeds can be expressed as functions of $j$,

\begin{equation}
(\bm{W}_L^*,\bm{W}_M^*,\bm{W}_R^*,S_P,S_C)
=\mathcal{R}(\bm{W}_L,\bm{W}_R,S_L,S_R,Q_{\mathrm{lat}},\sigma,\kappa,j).
\label{eq:eq51}
\end{equation}

\noindent Since the liquid-side and gas-side states adjacent to the phase interface are also functions of
$j$, substituting the reduced solution into the multi-species Schrage--Knudsen equation yields a
scalar nonlinear equation for the mass flux,

\begin{equation}
j=f_m\left(\bm{W}_{\mathrm{liq}}^*(j),\bm{W}_{\mathrm{gas}}^*(j)\right).
\label{eq:eq52}
\end{equation}

\noindent where $f_m$ denotes the right-hand side of the multi-species Schrage--Knudsen equation in
Eq.~\eqref{eq:eq47}. Equation~\eqref{eq:eq52} is solved by iteratively updating the interfacial
states and the mass flux until the change in $j$ is below a prescribed tolerance. After $j$ is
determined, all intermediate states and wave speeds are reconstructed from the reduced system.

After the multi-species phase-change Riemann problem is solved, the interfacial exchange fluxes
in cut cells are constructed from the star states adjacent to the phase interface. The
interfacial exchange flux for the liquid phase is

\begin{equation}
\hat{\bm{X}}_{\mathrm{liq}}=-\Delta\Gamma
\begin{bmatrix}
j\\
(jv_{\mathrm{liq}}^*+p_{\mathrm{liq}}^*)n_x\\
(jv_{\mathrm{liq}}^*+p_{\mathrm{liq}}^*)n_y\\
(jv_{\mathrm{liq}}^*+p_{\mathrm{liq}}^*)n_z\\
j\left(e_{pc}^*+\frac{1}{2}v_{\mathrm{liq}}^{*2}\right)
+p_{\mathrm{liq}}^*v_{\mathrm{liq}}^*
\end{bmatrix},
\label{eq:eq53}
\end{equation}

\noindent while the corresponding flux for the gas mixture is

\begin{equation}
\hat{\bm{X}}_{\mathrm{gas}}=\Delta\Gamma
\begin{bmatrix}
j\\
(jv_{\mathrm{gas}}^*+p_{\mathrm{gas}}^*)n_x\\
(jv_{\mathrm{gas}}^*+p_{\mathrm{gas}}^*)n_y\\
(jv_{\mathrm{gas}}^*+p_{\mathrm{gas}}^*)n_z\\
j\left(e_{pc}^*+\frac{1}{2}v_{\mathrm{gas}}^{*2}\right)
+p_{\mathrm{gas}}^*v_{\mathrm{gas}}^*\\
j\\
0\\
\vdots\\
0
\end{bmatrix},
\label{eq:eq54}
\end{equation}

\noindent where $e_{pc}^*$ is evaluated on each side according to
Eq.~\eqref{eq:eq42}. The first entry of Eq.~\eqref{eq:eq54} gives the net gas-mixture mass
exchange, and the final entries prescribe the species exchanges in Eq.~\eqref{eq:eq37}. Because
no separate species equation is solved on the
liquid side, conservation is first checked for the variables shared by the two phases, namely
mass, momentum, and total energy. Adding the corresponding hydrodynamic parts of the two
exchange fluxes gives

\begin{equation}
\hat{\bm{X}}_{\mathrm{gas}}^{\mathrm{hydro}}
+\hat{\bm{X}}_{\mathrm{liq}}^{\mathrm{hydro}}
=\Delta\Gamma
\begin{bmatrix}
0\\
-\sigma\kappa n_x\\
-\sigma\kappa n_y\\
-\sigma\kappa n_z\\
jQ_{\mathrm{lat}}-\sigma\kappa S_P
\end{bmatrix}.
\label{eq:eq55}
\end{equation}

\noindent The zero mass term confirms that the hydrodynamic exchange terms conserve total mass
across the interface exactly. The nonzero momentum and energy terms are physical
interfacial contributions associated with surface tension and latent heat, as discussed by Long
et al.~\cite{long2023conservativePhaseChange}. The species balance is checked separately through
the gas-side species equations. Since the liquid phase is composed only of the phase-changing
species, the liquid mass term and the gas-side condensable-vapor species term satisfy

\begin{equation}
\hat{X}_{\mathrm{liq}}^{\rho}
+\hat{X}_{\mathrm{gas}}^{\rho Y_1}
=-\Delta\Gamma j+\Delta\Gamma j=0,
\label{eq:eq56}
\end{equation}

\noindent which shows that the phase-changing material is transferred conservatively across the interface.
GFM-based phase-change methods impose the interface coupling through ghost states
\cite{houim2013ghostFluidPhaseChange,das2020shockVaporizationDroplets}. By contrast, the present
cut-cell method prescribes the net species exchange directly through Eq.~\eqref{eq:eq54}.
Therefore, no separate gas-side boundary condition for the species diffusion flux is required.
Equations~\eqref{eq:eq53}--\eqref{eq:eq56} provide the conservative interfacial exchange terms
used in the cut-cell finite-volume update. The accuracy, conservation property, and robustness of
the resulting method are examined in the following section.

\section{Validation and numerical experiments}
\label{sec:results}

In this section, the proposed method is validated and assessed through a series of numerical
examples. Unless otherwise specified, the inviscid fluxes and the level-set equation are
discretized using the fifth-order WENO scheme \cite{jiang1996weno}. For the detonation cases,
the inviscid fluxes are reconstructed using the second-order MUSCL scheme with the van Leer
limiter \cite{vanleer1979muscl} to improve robustness in the presence of strong discontinuities.
The viscous stresses, heat conduction, and species diffusion terms are evaluated using
fourth-order central finite-difference approximations. Time integration of the flow and
level-set equations is performed using a second-order strong-stability-preserving Runge--Kutta
scheme \cite{gottlieb1998sspRungeKutta}. The Courant-Friedrichs-Lewy (CFL) number is set to 0.2
for reactive-flow cases and 0.5 for all other cases. To improve computational efficiency, a
wavelet-based adaptive multi-resolution (MR) algorithm is employed
\cite{han2014adaptiveMultiresolution}.

For the axisymmetric shock-droplet cases in
Sec.~\ref{sec:axisymmetric-shock-al-droplets}, the governing
equations~\eqref{eq:eq1} are solved in cylindrical coordinates $(z,r)$ under the assumption of
rotational symmetry. The 2D planar operators are retained on the $(z,r)$ grid, and the
cylindrical metric terms are added as source terms. The inviscid contribution is

\begin{equation}
\bm{S}_{\mathrm{axi}}^{\mathrm{inv}}
=-\frac{1}{r}
\begin{bmatrix}
\rho u_r\\
\rho u_r u_z\\
\rho u_r^2\\
(\rho E+p)u_r\\
\rho u_r Y_1\\
\vdots\\
\rho u_r Y_{N-1}
\end{bmatrix},
\label{eq:axisymmetric_source}
\end{equation}

\noindent where $u_z$ and $u_r$ are the axial and radial velocity components. The viscous,
thermal, and species-diffusion contributions are

\begin{equation}
\bm{S}_{\mathrm{axi}}^{v}
=\frac{1}{r}
\begin{bmatrix}
0\\
\tau_{rz}\\
\tau_{rr}-\tau_{\theta\theta}\\
u_z\tau_{rz}+u_r\tau_{rr}-Q_r\\
-J_{r,1}\\
\vdots\\
-J_{r,N-1}
\end{bmatrix},
\label{eq:axisymmetric_viscous_source}
\end{equation}

\noindent with
\[
\tau_{\theta\theta}
=\frac{2}{3}\mu\left(2\frac{u_r}{r}
-\frac{\partial u_r}{\partial r}
-\frac{\partial u_z}{\partial z}\right).
\]
For the liquid phase, the species-diffusion entries in
Eq.~\eqref{eq:axisymmetric_viscous_source} are omitted. The source terms are integrated in each
Runge--Kutta substage, and the axis singularity is avoided by using a cell-centered radial grid
with symmetry imposed at $r=0$.

\subsection{1D test cases}
\label{sec:one-dimensional}

\subsubsection{Impulsive evaporation of liquid water into air}
\label{sec:impulsive-evaporation}

The first case considers the impulsive evaporation of a planar liquid-water surface into air,
following the benchmark problem of Houim and Kuo \cite{houim2013ghostFluidPhaseChange}. The
1D computational domain has a length of $4~\mathrm{mm}$, and the initial liquid--gas
interface is located at $x=0.25~\mathrm{mm}$. Both the liquid water and the gas are initially at
$p=1~\mathrm{atm}$ and $T=365~\mathrm{K}$. The gas phase is modeled as a three-species mixture
of $\mathrm{H_2O}$, $\mathrm{N_2}$, and $\mathrm{O_2}$, and is initially set to dry air, with
mole ratio $X_{\mathrm{H_2O}}:X_{\mathrm{N_2}}:X_{\mathrm{O_2}}=0:79:21$. The material and
interfacial properties of liquid water are $\mu=0.306~\mathrm{mPa\cdot s}$,
$\lambda=0.677~\mathrm{W/(m\cdot K)}$, and $\sigma=0.073~\mathrm{N/m}$. Non-reflecting boundary
conditions are imposed at both ends of the domain, which is discretized using 3200 uniform
cells.

\begin{figure}[tbp]
  \centering
  \includegraphics[width=\linewidth]{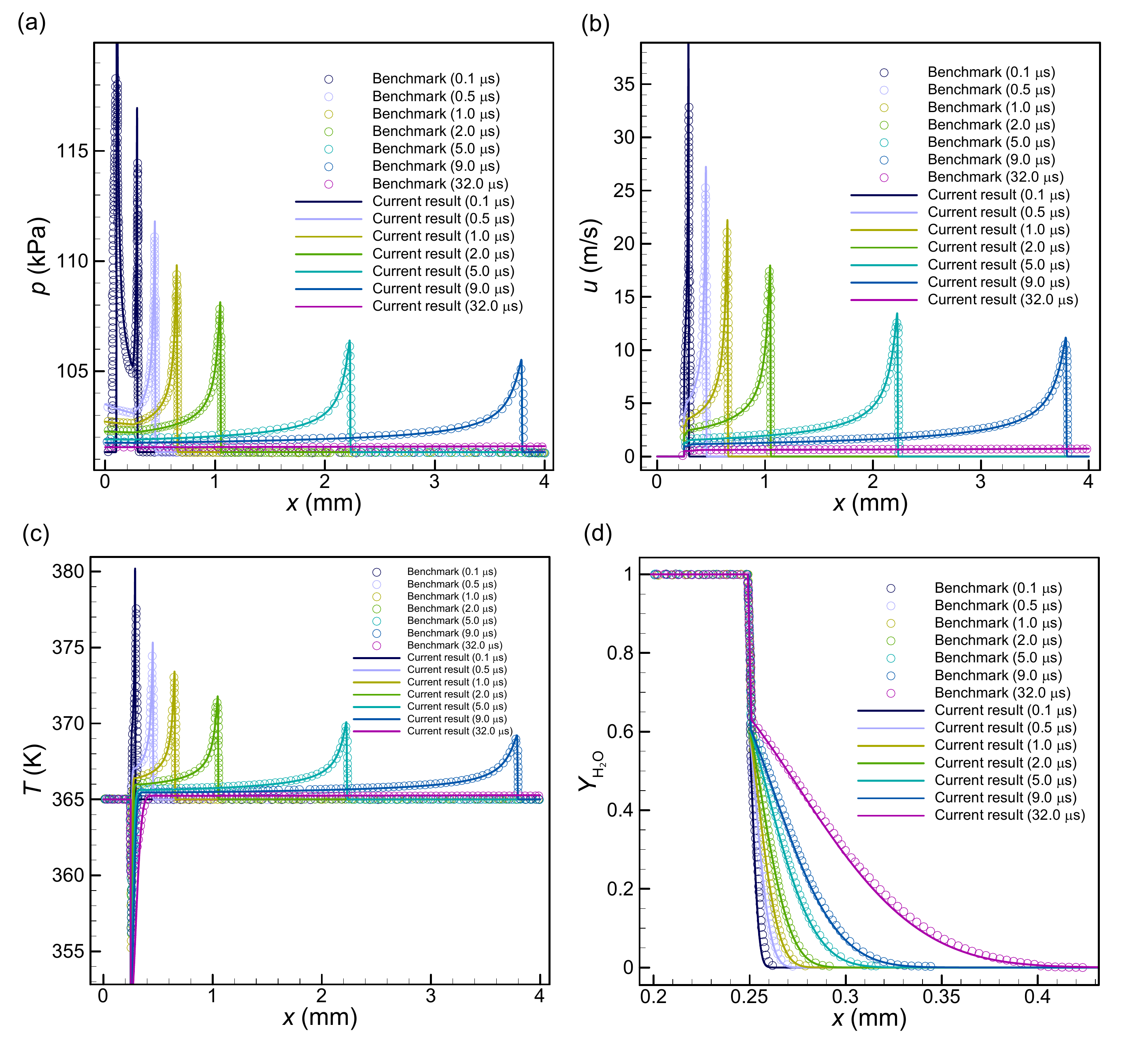}
  \caption{Computed profiles of (a) pressure, (b) velocity, (c) temperature, and (d) water mass
    fraction at various times for impulsive evaporation of liquid water into air. Present
    results (lines) are compared with the benchmark results (symbols) from Houim and Kuo
    \cite{houim2013ghostFluidPhaseChange}.}
  \label{fig:impulsive-evaporation-profiles}
\end{figure}

\begin{figure}[tbp]
  \centering
  \includegraphics[width=\linewidth]{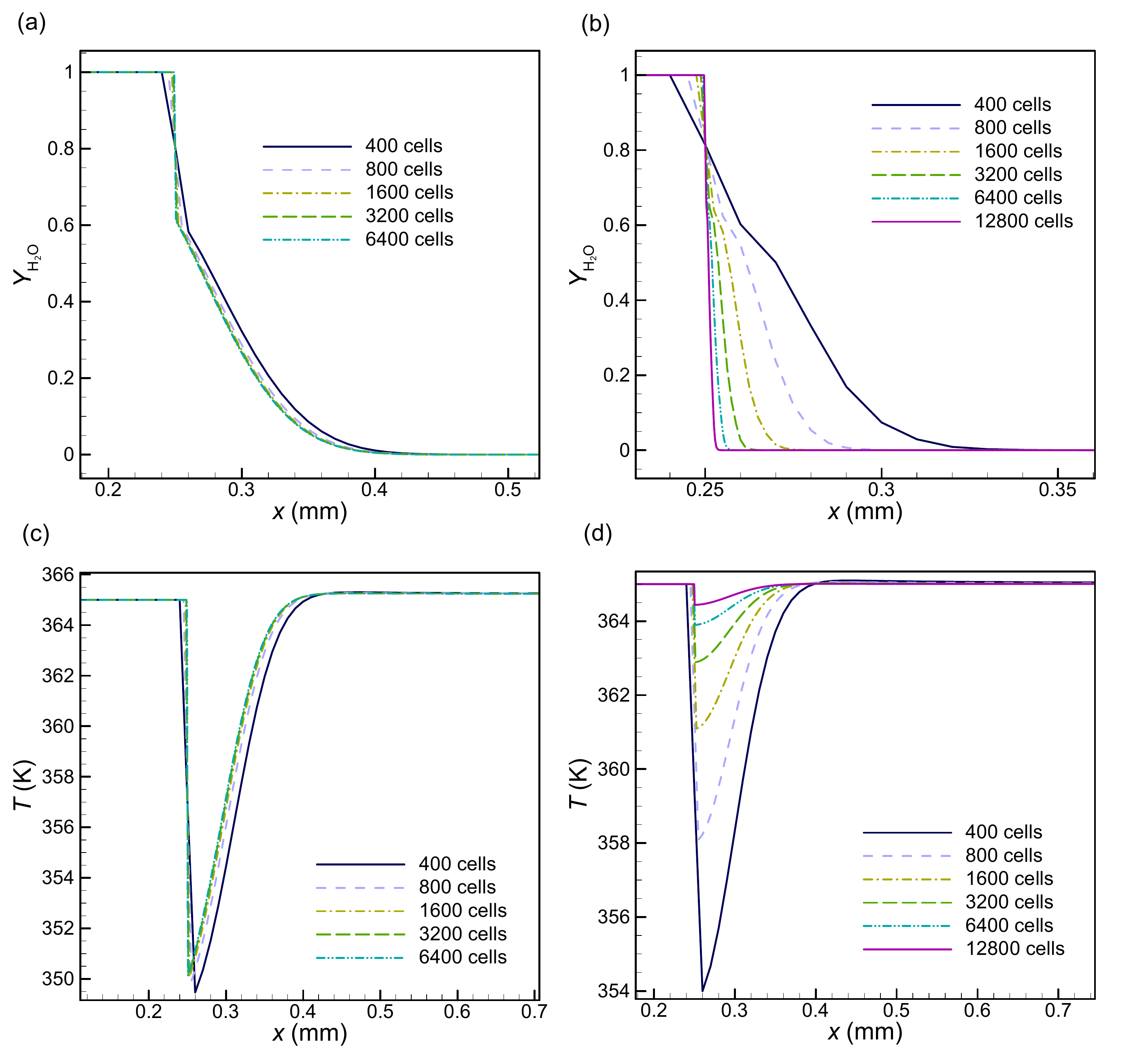}
  \caption{Grid convergence study for impulsive evaporation of liquid water into air: (a)--(b)
    water mass fraction and (c)--(d) temperature profiles. A comparison is made between the cases
    with diffusion (left panels, 400 to 6400 cells) and without diffusion (right panels,
    400 to 12800 cells).}
  \label{fig:evaporation-diffusion-grid}
\end{figure}

Fig.~\ref{fig:impulsive-evaporation-profiles} compares the computed profiles with the benchmark results of Houim and Kuo
\cite{houim2013ghostFluidPhaseChange}. The present solution agrees well with the benchmark,
confirming that the coupled effects of phase change and vapor diffusion are accurately captured.
For multi-species phase change, vapor diffusion controls the interfacial vapor distribution and
sustains the concentration gradient required for continued evaporation
\cite{houim2013ghostFluidPhaseChange}. To isolate this effect, additional calculations are
performed with and without vapor diffusion. As shown in
Fig.~\ref{fig:evaporation-diffusion-grid}, when diffusion is omitted, vapor generated at the
interface cannot be transported into the surrounding gas mixture by a physical mechanism. Grid
refinement then reduces numerical diffusion, causing the vapor to remain increasingly confined
to a thin layer adjacent to the liquid surface. The near-interface vapor concentration rapidly
approaches its saturation value, which weakens the vapor-concentration gradient and markedly
reduces the evaporation rate. In the grid-refinement limit, evaporation would therefore be
artificially suppressed in the absence of physical vapor transport. These results show that mass
diffusion is essential for grid-convergent and physically consistent simulations of
multi-species phase change.

\FloatBarrier
\subsubsection{Impulsive condensation of water vapor from air}
\label{sec:impulsive-condensation}

\begin{figure}[tbp]
  \centering
  \includegraphics[width=\linewidth]{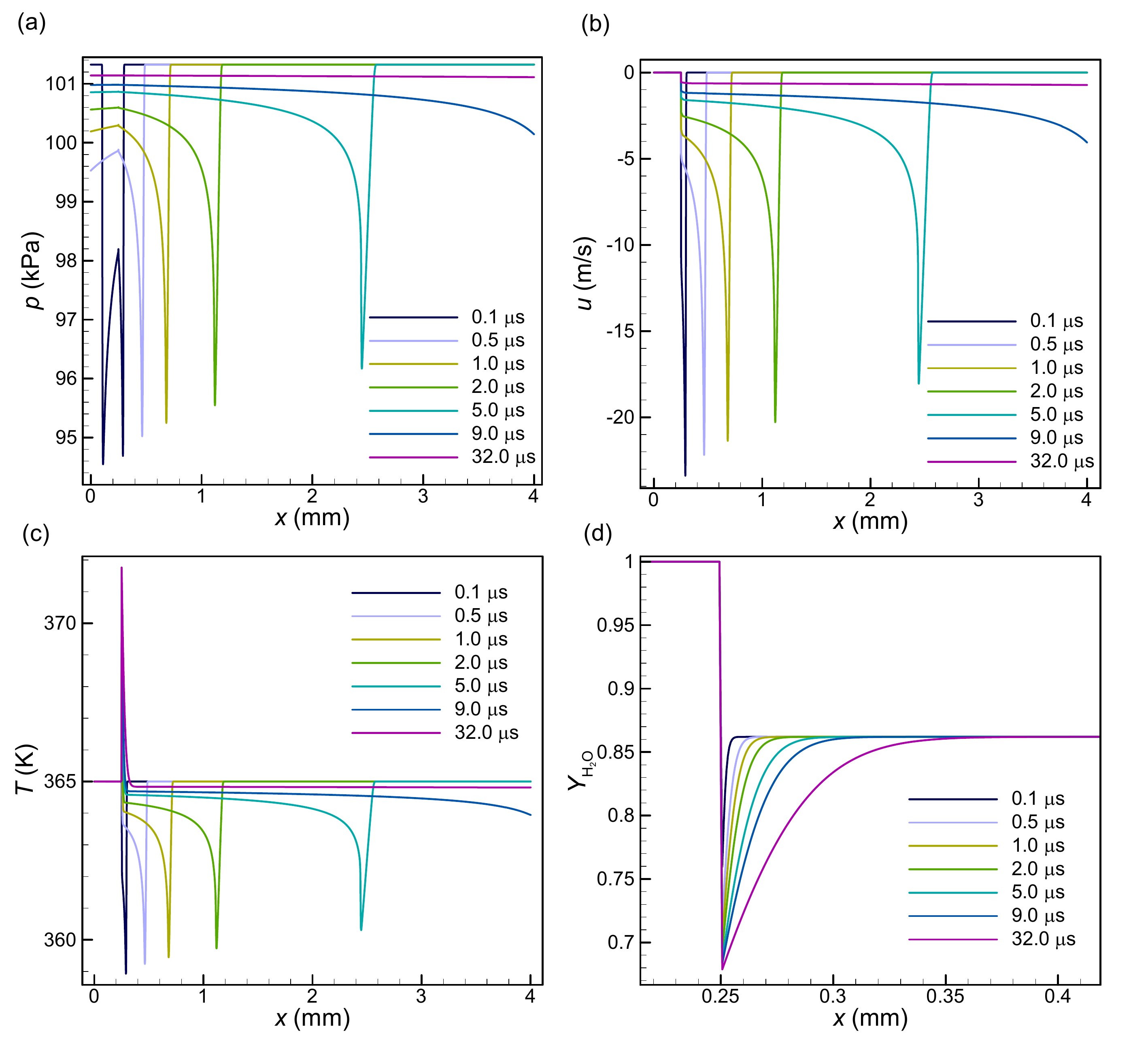}
  \caption{Computed profiles of (a) pressure, (b) velocity, (c) temperature, and (d) water mass
    fraction at various times for impulsive condensation of water vapor from air.}
  \label{fig:impulsive-condensation-profiles}
\end{figure}

Here, we consider the condensation of water vapor from a gas mixture to assess the capability of
the present method for multi-species condensation.
The computational domain and initial conditions are identical to those of the
evaporation case in Sec.~\ref{sec:impulsive-evaporation}, except for the initial gas
composition. To induce condensation, the gas mixture is initialized with the mole ratio
$X_{\mathrm{H_2O}}:X_{\mathrm{N_2}}:X_{\mathrm{O_2}}=1000:79:21$, corresponding to a
supersaturated water-vapor state. The profiles in
Fig.~\ref{fig:impulsive-condensation-profiles} exhibits trends opposite to those in evaporation.
Vapor removal at the liquid surface creates a local mass sink, drives the gas velocity toward
the liquid, and produces an expansion-type pressure depression. Latent-heat release yields a
localized temperature rise, while vapor consumption and diffusion produce a gradual decrease in
water-vapor mass fraction from the far field to the interface.

\begin{figure}[tbp]
  \centering
  \includegraphics[width=\linewidth]{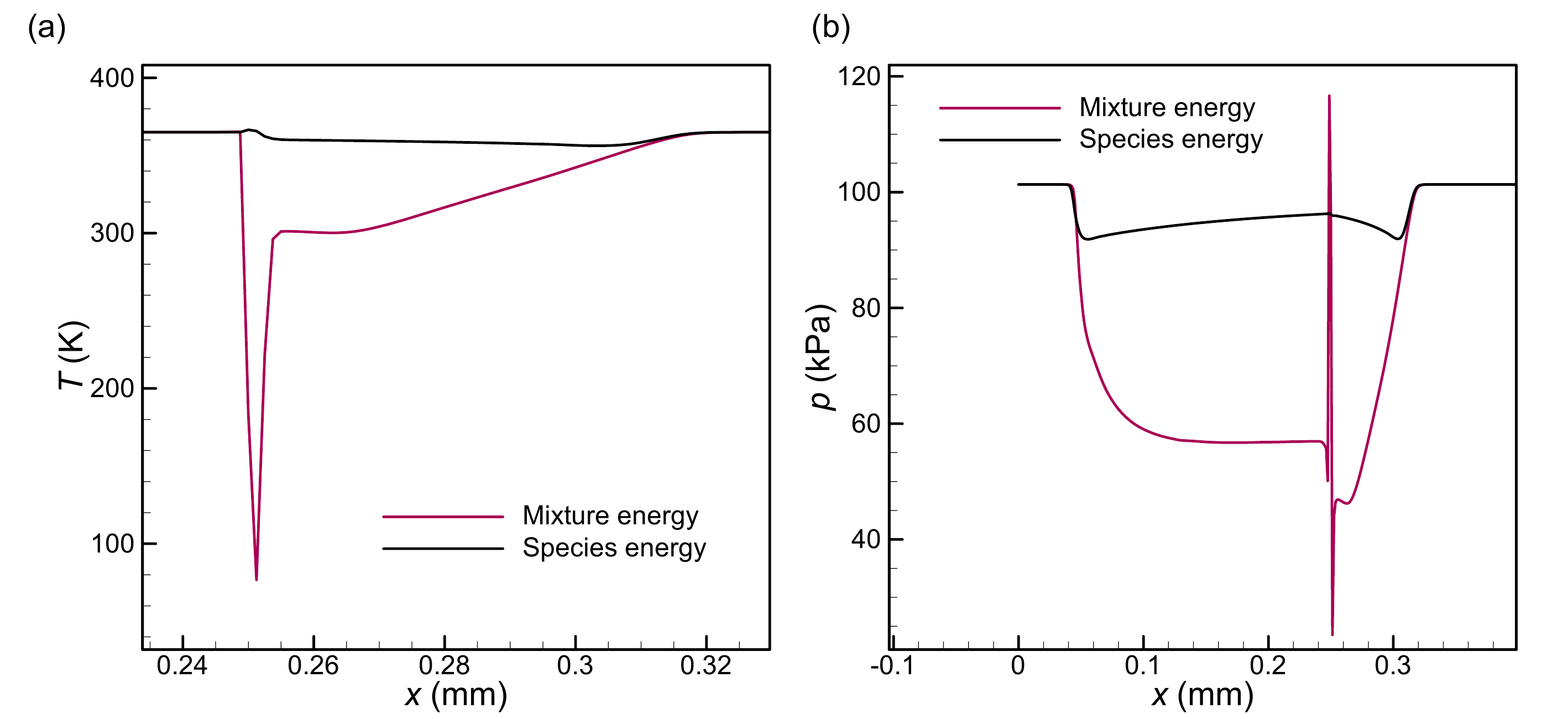}
  \caption{Comparison of interfacial energy-jump conditions based on the phase-change species
    energy and the mixture energy for multi-species phase change: (a) temperature and (b)
    pressure profiles.}
  \label{fig:condensation-energy-jump-comparison}
\end{figure}

The condensation case provides a stringent test of the interfacial energy jump condition in
Eq.~\eqref{eq:eq43}, because the mass flux is directed from the multi-species gas phase into
the liquid. If the gas-mixture internal energy is used, the energy carried by non-condensable
species is incorrectly included in the condensation flux. This error is weak in evaporation,
where the newly generated interfacial gas layer $\bm{W}_M^*$ is nearly pure vapor, but it
becomes pronounced in condensation because the gas-side state contains substantial
non-condensable species. To isolate this effect, a diagnostic condensation calculation is
performed with a prescribed mass flux of $j=-10~\mathrm{kg/(m^2\cdot s)}$, where the negative
sign denotes transfer from the gas phase to the liquid phase. As shown in
Fig.~\ref{fig:condensation-energy-jump-comparison}, using the mixture internal energy violates
the species-selective construction of $e_{pc}$ in Eq.~\eqref{eq:eq42} and the energy jump
condition in Eq.~\eqref{eq:eq43}. The resulting fictitious energy transport distorts the
interfacial energy balance, first producing a spurious temperature depression and then, through
the equation of state and jump conditions, a sharp pressure spike. These artifacts disappear when the
phase-change energy flux is constructed from
the internal energy of the condensable vapor species alone. Thus, for multi-species phase
change, the interfacial energy jump condition must use the internal energy of the
phase-changing species rather than the gas-mixture internal energy.

\FloatBarrier
\subsubsection{Vaporization of an aluminum slab with chemical reactions}
\label{sec:aluminum-slab}

In this case, we consider the vaporization of a planar aluminum slab into air with chemical
reactions, following the configuration of Houim and Kuo \cite{houim2013ghostFluidPhaseChange}.
This problem provides a coupled test of phase change, multi-species transport, and chemical
reactions. The computational domain and initial interface location are the same as those in
Sec.~\ref{sec:impulsive-evaporation}. The liquid aluminum and gas phases are initialized at
$T=2750~\mathrm{K}$ and $300~\mathrm{K}$, respectively. The reacting gas mixture contains nine
species: $\mathrm{Al}$, $\mathrm{O_2}$, $\mathrm{O}$,
$\mathrm{AlO}$, $\mathrm{AlO_2}$, $\mathrm{Al_2O}$, $\mathrm{Al_2O_2}$, $\mathrm{Al_2O_3(l)}$,
and $\mathrm{N_2}$. The gas-phase portion of the domain is initially pure air. Following Ref.
\cite{houim2013ghostFluidPhaseChange}, the aluminum oxidation mechanism includes nine species
and nine reactions. The reaction rate follows the Arrhenius form of Eq.~\eqref{eq:eq21}, and the
complete reaction mechanism and rate parameters are taken from Ref.~\cite{houim2013ghostFluidPhaseChange}.
The material and
interfacial properties of liquid aluminum are $\mu=0.362~\mathrm{mPa\cdot s}$,
$\lambda=196~\mathrm{W/(m\cdot K)}$, and $\sigma=0.526~\mathrm{N/m}$.
As in the reduced mechanism of Houim and Kuo~\cite{houim2013ghostFluidPhaseChange},
the label ``$(l)$'' in $\mathrm{Al_2O_3(l)}$ identifies the thermochemical state assigned to the
oxidation product. It is transported as a species of the single gas mixture and therefore uses the
mixture EOS and gas-phase transport model in Eqs.~\eqref{eq:eq14}--\eqref{eq:eq15}. It does not
represent a separately resolved condensed-alumina phase with an independent volume, momentum, or
deposition model.

\begin{figure}[tbp]
  \centering
  \includegraphics[width=\linewidth]{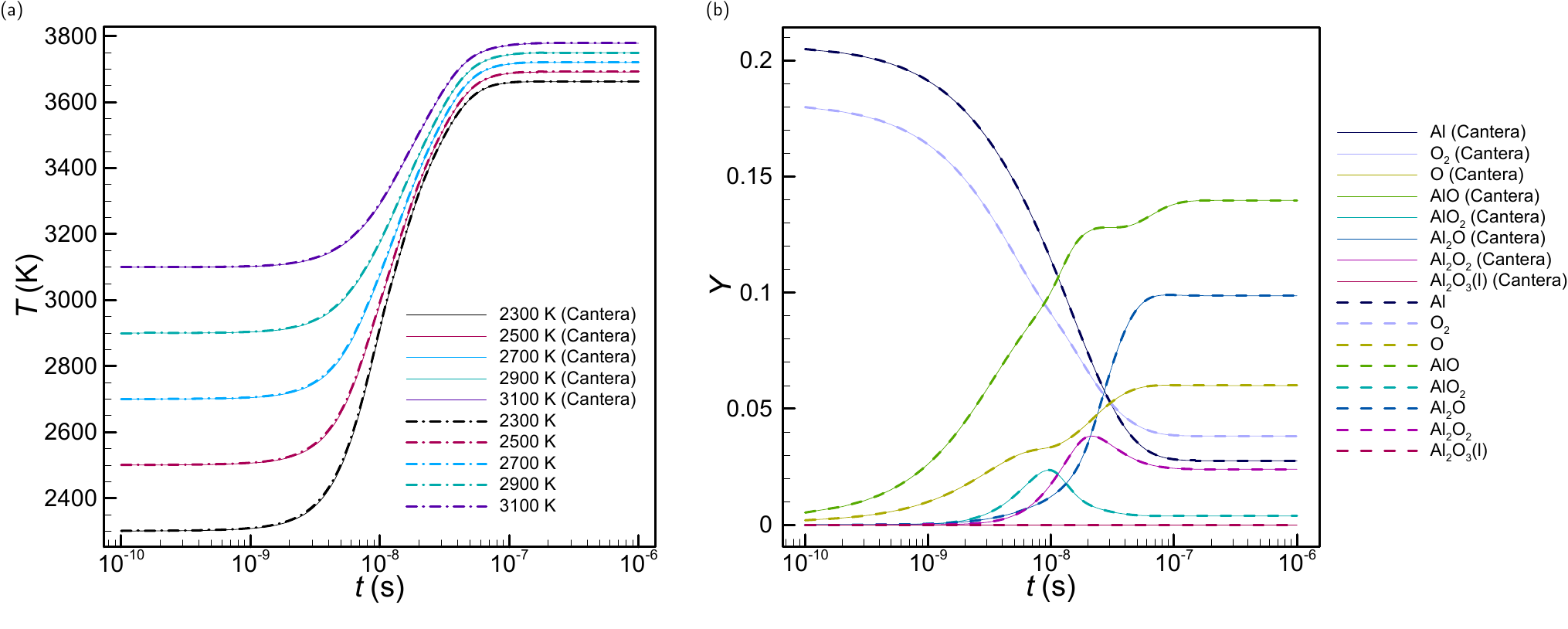}
  \caption{Code-to-code verification of the aluminum oxidation kinetics against Cantera:
    (a) temporal evolution of temperature for various initial temperatures and (b) temporal
    evolution of species mass fractions at $T_0=2300~\mathrm{K}$.}
  \label{fig:al-reactor-validation}
\end{figure}

The implementation of the homogeneous aluminum oxidation kinetics is first verified against
Cantera calculations in a constant-volume, adiabatic reactor using identical initial states.
Fig.~\ref{fig:al-reactor-validation}(a)
compares the temporal evolution of temperature for several initial temperatures. The present
results agree well with the Cantera solutions, reproducing the ignition delay, the rapid
temperature rise, and the final reactor temperature. Fig.~\ref{fig:al-reactor-validation}(b)
compares the species mass-fraction histories at $T_0=2300~\mathrm{K}$. The consumption of
$\mathrm{Al}$ and $\mathrm{O_2}$ and the formation of aluminum oxide intermediates and products
are reproduced. This agreement verifies the implementation of the reaction mechanism.

\begin{figure}[tbp]
  \centering
  \includegraphics[width=\linewidth]{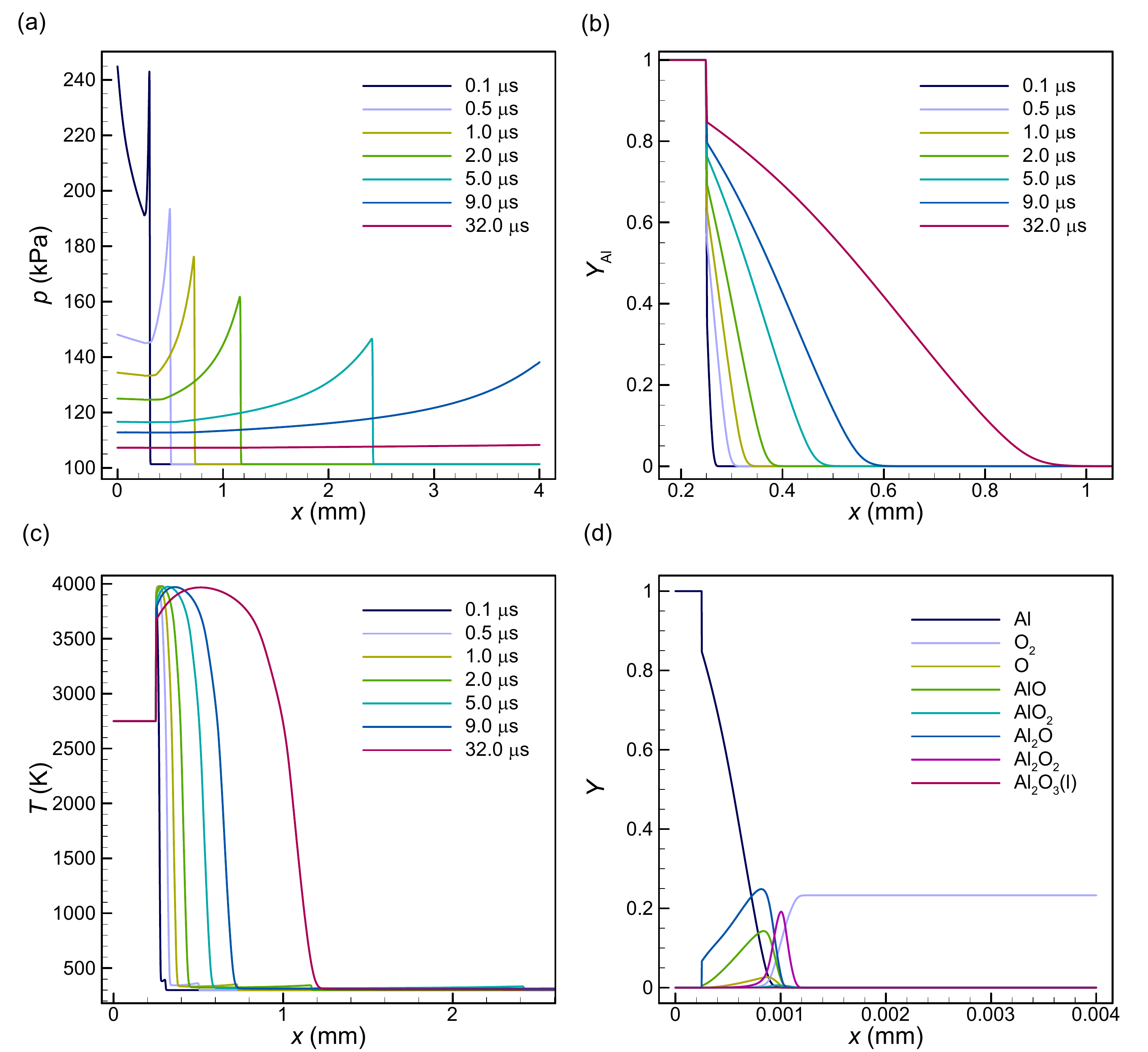}
  \caption{Computed spatial profiles for the vaporization of an aluminum slab with chemical
    reactions: (a) pressure, (b) Al mass fraction, and (c) temperature at various times, and (d)
    spatial distribution of the species mass fractions at $t=32.0~\mu\mathrm{s}$.}
  \label{fig:al-slab-reaction-profiles}
\end{figure}

The verified kinetics implementation is then used in the 1D aluminum-slab vaporization
case to test the coupling among phase change, species diffusion, and finite-rate chemistry.
Fig.~\ref{fig:al-slab-reaction-profiles} shows the resulting spatial profiles.
Impulsive aluminum vaporization generates pressure waves that propagate into the gas phase,
analogous to the water-evaporation case in Sec.~\ref{sec:impulsive-evaporation}.
Meanwhile, Al vapor is supplied from the liquid surface and transported into the initially
quiescent gas region. As the vapor
mixes with oxygen, gas-phase oxidation occurs and releases heat, producing a high-temperature
reaction layer adjacent to the vaporizing surface. The Al mass fraction profiles show the
progressive expansion of the aluminum-vapor layer, while the temperature profiles indicate the
development of the reacting thermal layer. At $t=32.0~\mu\mathrm{s}$, the species profiles in
Fig.~\ref{fig:al-slab-reaction-profiles}(d) reveals
the structure of the reacting diffusion layer. $\mathrm{Al}$ and $\mathrm{O_2}$ are consumed,
while $\mathrm{O}$, $\mathrm{AlO}$, $\mathrm{AlO_2}$, $\mathrm{Al_2O}$, $\mathrm{Al_2O_2}$, and
$\mathrm{Al_2O_3(l)}$ are formed. The coexistence of reactants, intermediates, and products over
a finite spatial region indicates that the flame structure is governed by coupled convection,
diffusion, and finite-rate chemistry.

\FloatBarrier
\subsection{2D water-droplet evaporation and condensation}
\label{sec:2d-water-droplet-tests}

\subsubsection{Impulsive evaporation of a circular water droplet}
\label{sec:2d-water-evaporation}

\begin{figure}[tbp]
  \centering
  \includegraphics[width=\linewidth]{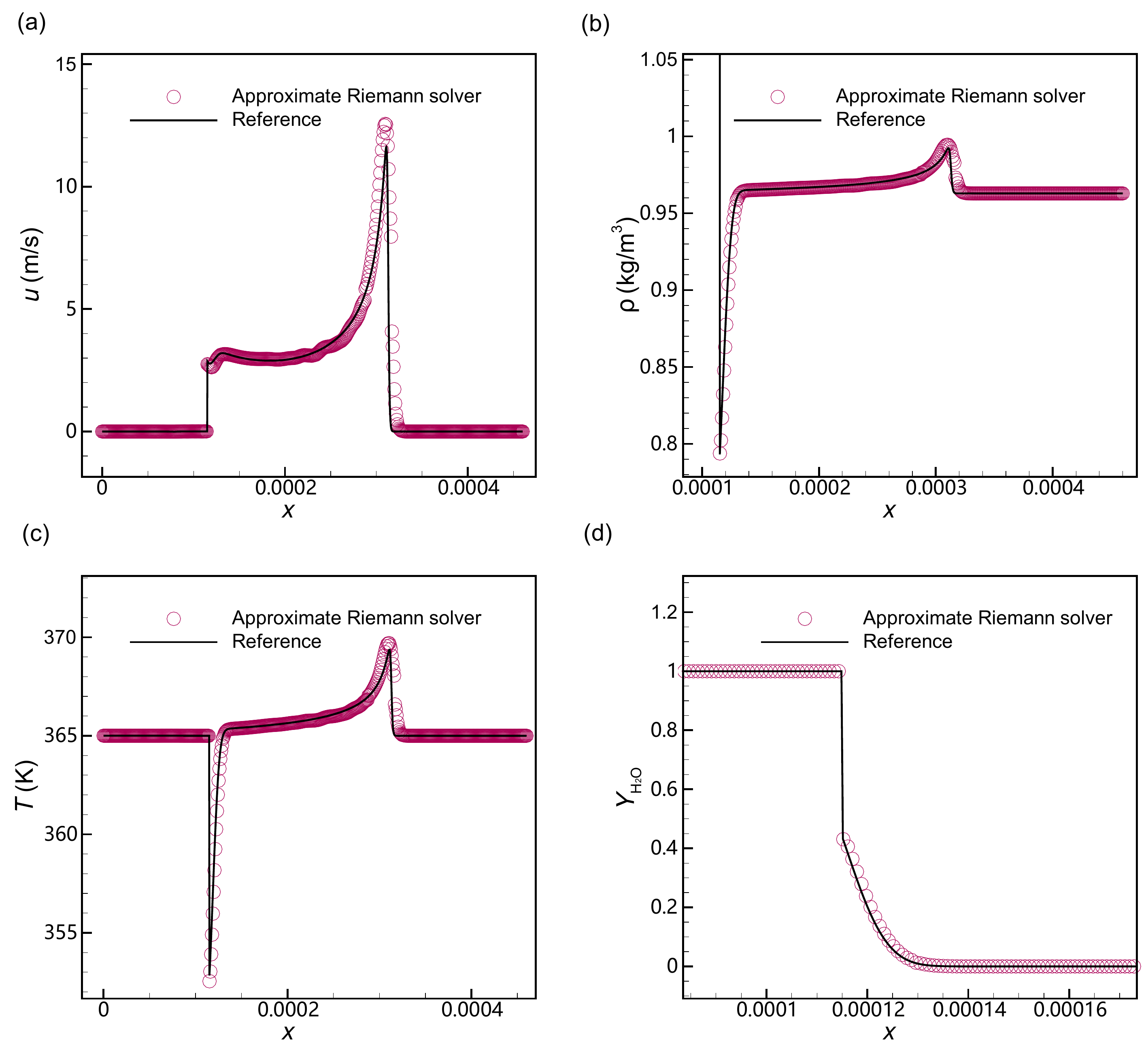}
  \caption{Impulsive evaporation of a circular water droplet: (a) the velocity profile, (b) the
    density profile, (c) the temperature profile, and (d) the $Y_{\mathrm{H_2O}}$ mass fraction
    profile. The simulation is carried out on an MR grid with an effective resolution of
    $512\times512$. The results shown here are extracted along the
    $x$-axis. With 3200 grid points, the numerical reference solution is obtained by employing the radially
    symmetric 1D approach
    \cite{long2023conservativePhaseChange,toro2009riemannSolvers}.}
  \label{fig:2d-water-evaporation-profiles}
\end{figure}

We next consider impulsive evaporation from a 2D circular water droplet into air. A square
domain of side length $2D_0$ is used, where $D_0=230~\mu\mathrm{m}$. The droplet center is
placed at the origin, and only one-quarter of the configuration is computed with symmetry
conditions on the coordinate axes. The calculation is performed on an MR grid with an effective
resolution of $512\times512$. The initial pressure, gas temperature, gas composition, and liquid
temperature are the same as those in the 1D evaporation case in
Sec.~\ref{sec:impulsive-evaporation}. A radially symmetric 1D calculation with 3200 grid points
is used as a numerical reference solution
\cite{long2023conservativePhaseChange,toro2009riemannSolvers}.
Fig.~\ref{fig:2d-water-evaporation-profiles} compares the velocity, density, temperature, and
$Y_{\mathrm{H_2O}}$ profiles extracted along the $x$-axis at $t=0.5~\mu\mathrm{s}$. The 2D
solution on the $512\times512$ MR grid agrees closely with the radially symmetric numerical reference
solution. Compared with the planar 1D case in Sec.~\ref{sec:impulsive-evaporation}, radial
spreading reduces the pressure, velocity, and temperature perturbations and accelerates the decay
of vapor concentration away from the interface. The agreement supports the consistency of the
multidimensional cut-cell implementation for phase change at a curved interface.

\FloatBarrier
\subsubsection{Impulsive condensation on a circular water droplet}
\label{sec:2d-water-condensation}

\begin{figure}[tbp]
  \centering
  \includegraphics[width=\linewidth]{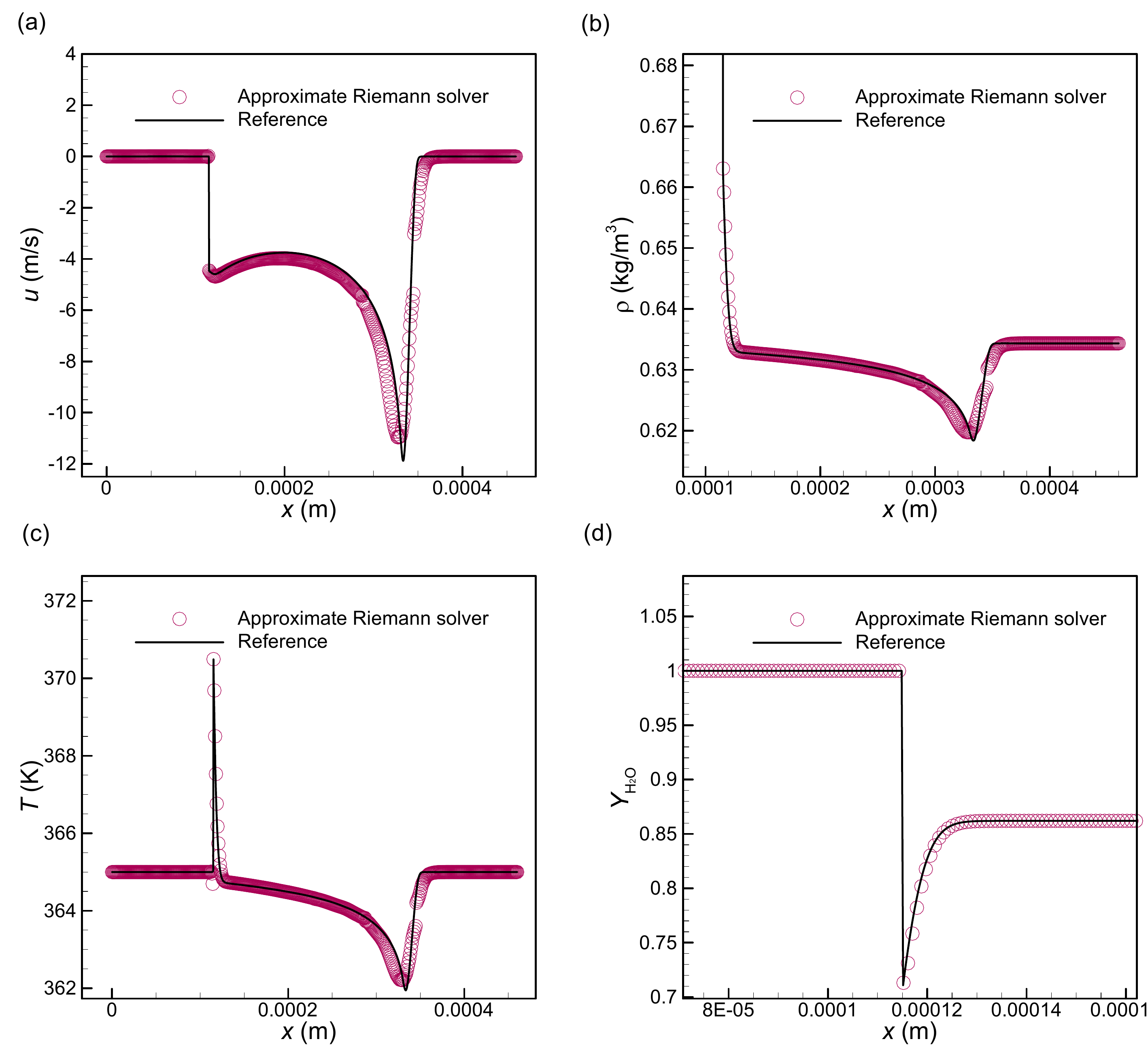}
  \caption{Impulsive condensation on a circular water droplet: (a) the velocity
    profile, (b) the density profile, (c) the temperature profile, and (d) the
    $Y_{\mathrm{H_2O}}$ mass fraction profile. The simulation is carried out on an MR grid with
    an effective resolution of $512\times512$. The results shown here are extracted along the
    $x$-axis. With 3200 grid points, the numerical reference solution is obtained by employing the radially
    symmetric 1D approach
    \cite{long2023conservativePhaseChange,toro2009riemannSolvers}.}
  \label{fig:2d-water-condensation-profiles}
\end{figure}

We next use the same circular-droplet geometry to examine condensation. The domain, boundary
conditions, and $512\times512$ effective MR resolution are unchanged from
Sec.~\ref{sec:2d-water-evaporation}. Condensation is induced by replacing the dry-air state with
the supersaturated gas mixture used in Sec.~\ref{sec:impulsive-condensation}, while keeping the
same initial pressure, gas temperature, and liquid temperature. A radially symmetric 1D
calculation with 3200 grid points is used as a numerical reference solution
\cite{long2023conservativePhaseChange,toro2009riemannSolvers}.
Fig.~\ref{fig:2d-water-condensation-profiles} compares the velocity, density, temperature, and
$Y_{\mathrm{H_2O}}$ profiles extracted along the $x$-axis at $t=0.5~\mu\mathrm{s}$ with the
numerical reference solution. The agreement supports the consistency of the multidimensional
interfacial treatment when the phase-change mass flux is reversed and vapor is removed from the
gas mixture.

To assess conservation, we monitor the global mass and energy budgets during the 2D droplet
simulations. Following Long et al.~\cite{long2023conservativePhaseChange}, the common cell volume
factor is omitted. With $\chi_{\mathrm{gas},i}=\alpha_i$ and
$\chi_{\mathrm{liq},i}=1-\alpha_i$, where $\alpha_i$ is the gas volume fraction in cell $i$, the
phase budgets are

\begin{equation}
M_q^n=\sum_i\chi_{q,i}^n\rho_{q,i}^n,\qquad
E_q^n=\sum_i\chi_{q,i}^n\rho_{q,i}^nE_{q,i}^n,\qquad q\in\{\mathrm{gas},\mathrm{liq}\}.
\label{eq:phase_budgets}
\end{equation}

\noindent The total mass and energy are then defined by

\begin{equation}
\begin{aligned}
M_{\mathrm{total}}^n&=M_{\mathrm{gas}}^n+M_{\mathrm{liq}}^n,\\
E_{\mathrm{total}}^n&=E_{\mathrm{gas}}^n+E_{\mathrm{liq}}^n
+\left(M_{\mathrm{liq}}^n-M_{\mathrm{liq}}^0\right)Q_{\mathrm{lat}}.
\end{aligned}
\label{eq:total_energy}
\end{equation}

\noindent The superscripts $0$ and $n$ denote the initial state and the $n$-th time step,
respectively. The final term in Eq.~\eqref{eq:total_energy} accounts for the latent heat carried
by the phase-changing mass and balances the $jQ_{\mathrm{lat}}$ interfacial energy exchange in
Eq.~\eqref{eq:eq55}. The plotted normalized phase variations and global residuals are

\begin{equation}
\begin{aligned}
\Delta M_q^n&=\frac{M_q^n-M_q^0}{M_{\mathrm{total}}^0}, &
\Delta E_q^n&=\frac{E_q^n-E_q^0}{E_{\mathrm{total}}^0},
&&q\in\{\mathrm{gas},\mathrm{liq}\},\\
\Delta M_{\mathrm{total}}^n&=\frac{M_{\mathrm{total}}^n-M_{\mathrm{total}}^0}{M_{\mathrm{total}}^0}, &
\Delta E_{\mathrm{total}}^n&=\frac{E_{\mathrm{total}}^n-E_{\mathrm{total}}^0}{E_{\mathrm{total}}^0}.
\end{aligned}
\label{eq:conservation_variations}
\end{equation}

\begin{figure}[tbp]
  \centering
  \includegraphics[width=\linewidth]{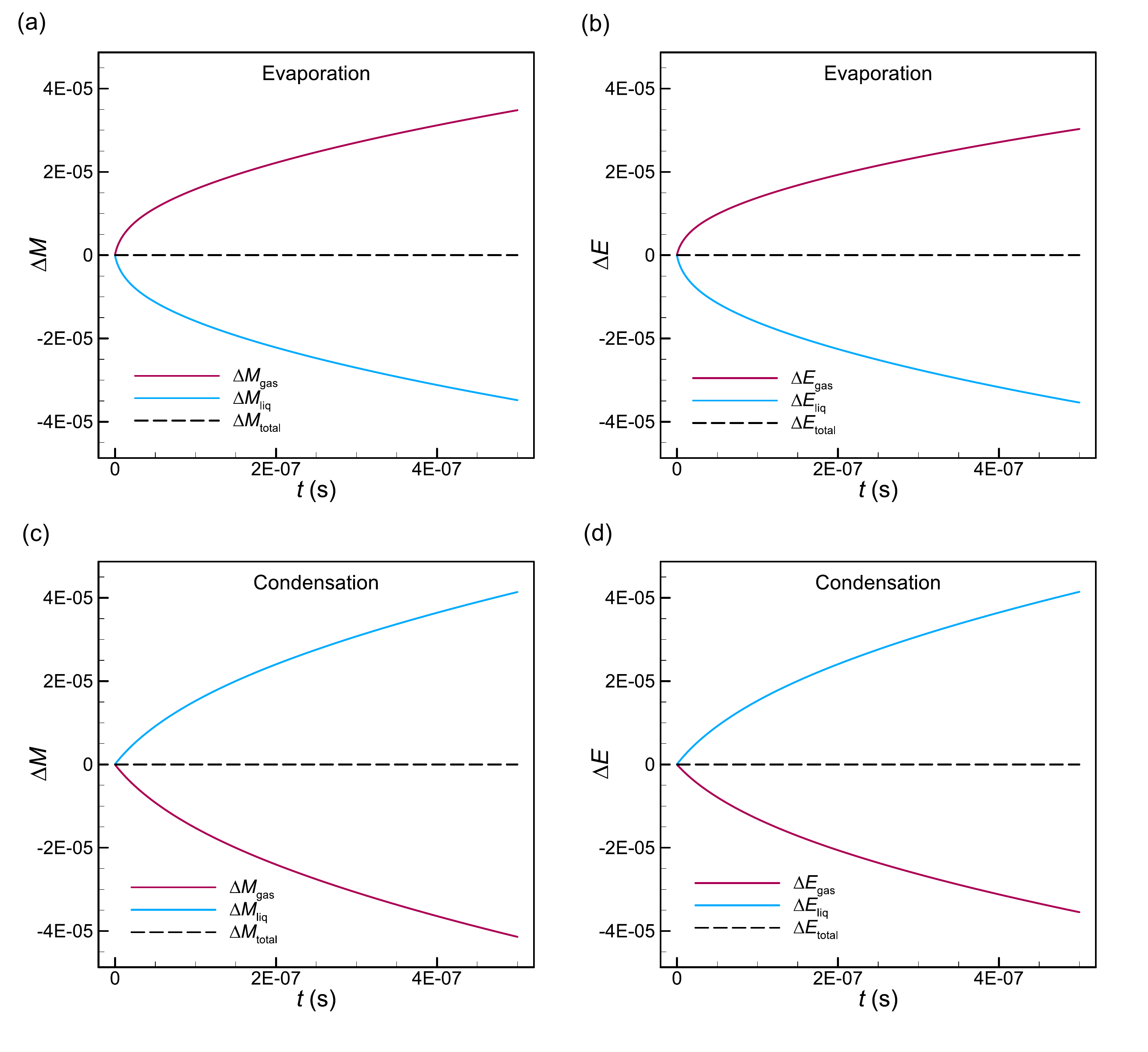}
  \caption{Conservation of total mass and augmented energy for 2D water droplet evaporation and condensation:
    (a)--(b) evaporation and (c)--(d) condensation.}
  \label{fig:2d-water-conservation}
\end{figure}

\noindent For evaporation, $\Delta M_{\mathrm{gas}}>0$ and $\Delta M_{\mathrm{liq}}<0$, with
reversed signs otherwise. Fig.~\ref{fig:2d-water-conservation} shows the normalized mass and energy
variations of the gas and liquid phases, together with the residuals of total mass and total
energy, for 2D droplet evaporation and condensation.

During evaporation, gas-phase mass and energy increase as their liquid-phase counterparts
decrease. The signs reverse during condensation. The total energy defined in
Eq.~\eqref{eq:total_energy} consistently accounts for the energy carried by the phase-changing
mass, so the total-mass and total-energy curves remain indistinguishable from zero on the scale
of Fig.~\ref{fig:2d-water-conservation}, demonstrating global conservation by the present
cut-cell interfacial coupling.

\FloatBarrier
\subsection{Axisymmetric shock interaction with aluminum droplets}
\label{sec:axisymmetric-shock-al-droplets}

\subsubsection{Shock interaction with a vaporizing aluminum droplet}
\label{sec:2d-al-droplet-evaporation}

\begin{figure}[!htbp]
  \centering
  \includegraphics[width=0.92\linewidth]{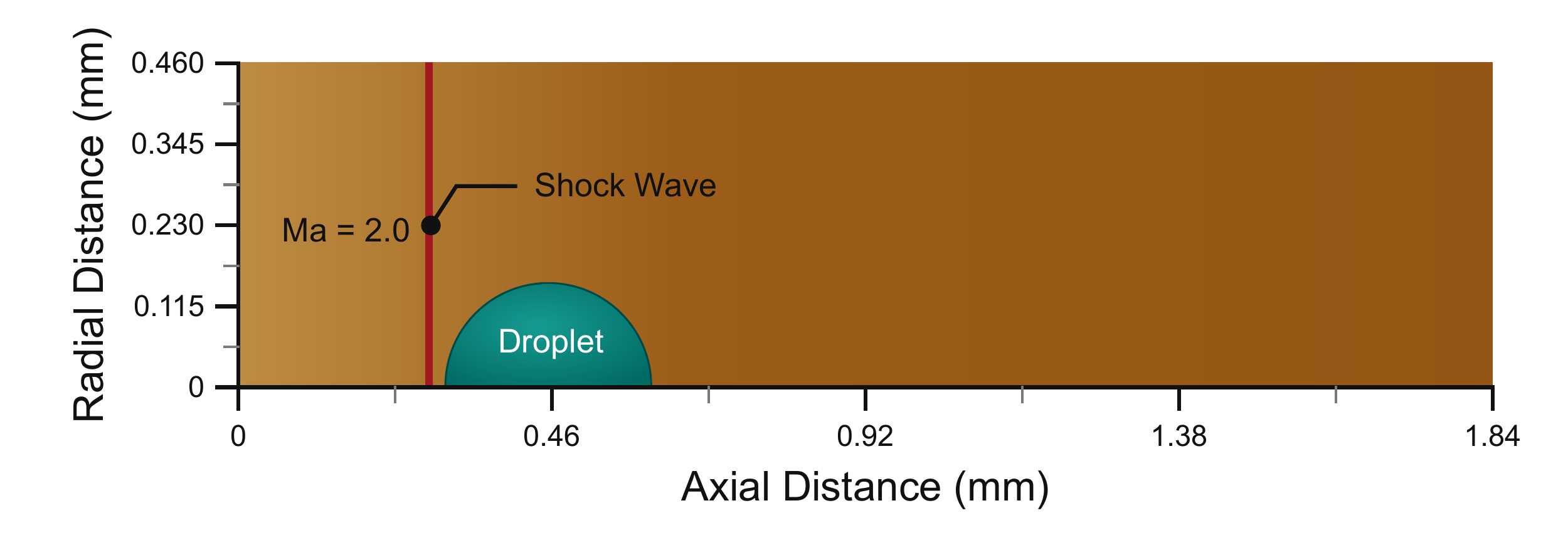}
  \caption{Schematic of the axisymmetric computational domain for a shock wave
  interacting with an aluminum droplet.}
  \label{fig:shock-al-droplet-domain}
\end{figure}

Next, we consider the interaction between a shock wave and a vaporizing aluminum droplet to
validate the present method for strongly compressible phase-change flows. The axisymmetric
computational domain is shown in Fig.~\ref{fig:shock-al-droplet-domain}. The axial and radial
extents are $8D_0$ and $2D_0$, respectively, where $D_0=230~\mu\mathrm{m}$ is the initial
droplet diameter. The liquid aluminum droplet is placed on the symmetry axis with its center at
$z=2D_0$, and the incident planar shock is initially located at $z=1.4D_0$. The shock Mach
number is $M_s=2.0$, and the shock propagates in the positive axial direction. The gas phase is
initialized as air, with oxygen and nitrogen mole fractions of 0.21 and 0.79, respectively. The
material properties of liquid aluminum and the thermodynamic and transport properties of
aluminum vapor are taken from Das and Udaykumar~\cite{das2020shockVaporizationDroplets}. The initial
pre-shock, post-shock, and droplet states are listed in
Table~\ref{tab:shock-al-initial-states}. An inflow condition is imposed at the left axial
boundary, outflow conditions are imposed at the right axial and outer radial boundaries, and
symmetry is imposed on the axis. A six-level MR adaptive grid is used, giving an effective
resolution of $1024\times4096$ in the radial and axial directions. This corresponds to 512 grid
cells across the initial droplet diameter, slightly higher than the 460 cells per diameter used
by Das and Udaykumar~\cite{das2020shockVaporizationDroplets}.

\begin{table}[!htbp]
  \centering
  \caption{Initial states for the Mach 2 shock interaction with an
  vaporizing aluminum droplet.}
  \label{tab:shock-al-initial-states}
  \begin{tabular}{lccc}
    \hline
    Variable & Pre-shock gas & Post-shock gas & Droplet \\
    \hline
    Density, $\rho$ ($\mathrm{kg/m^3}$) & 1.172 & 3.127 & 2003.0 \\
    Axial velocity, $u_z$ ($\mathrm{m/s}$) & 0.0 & 434.905 & 0.0 \\
    Pressure, $p$ ($\mathrm{Pa}$) & 101325.0 & 455880.360 & 101325.0 \\
    Temperature, $T$ ($\mathrm{K}$) & 300.0 & 505.885 & 2750.0 \\
    \hline
  \end{tabular}
\end{table}

\begin{figure}[tbp]
  \centering
  \includegraphics[width=\linewidth]{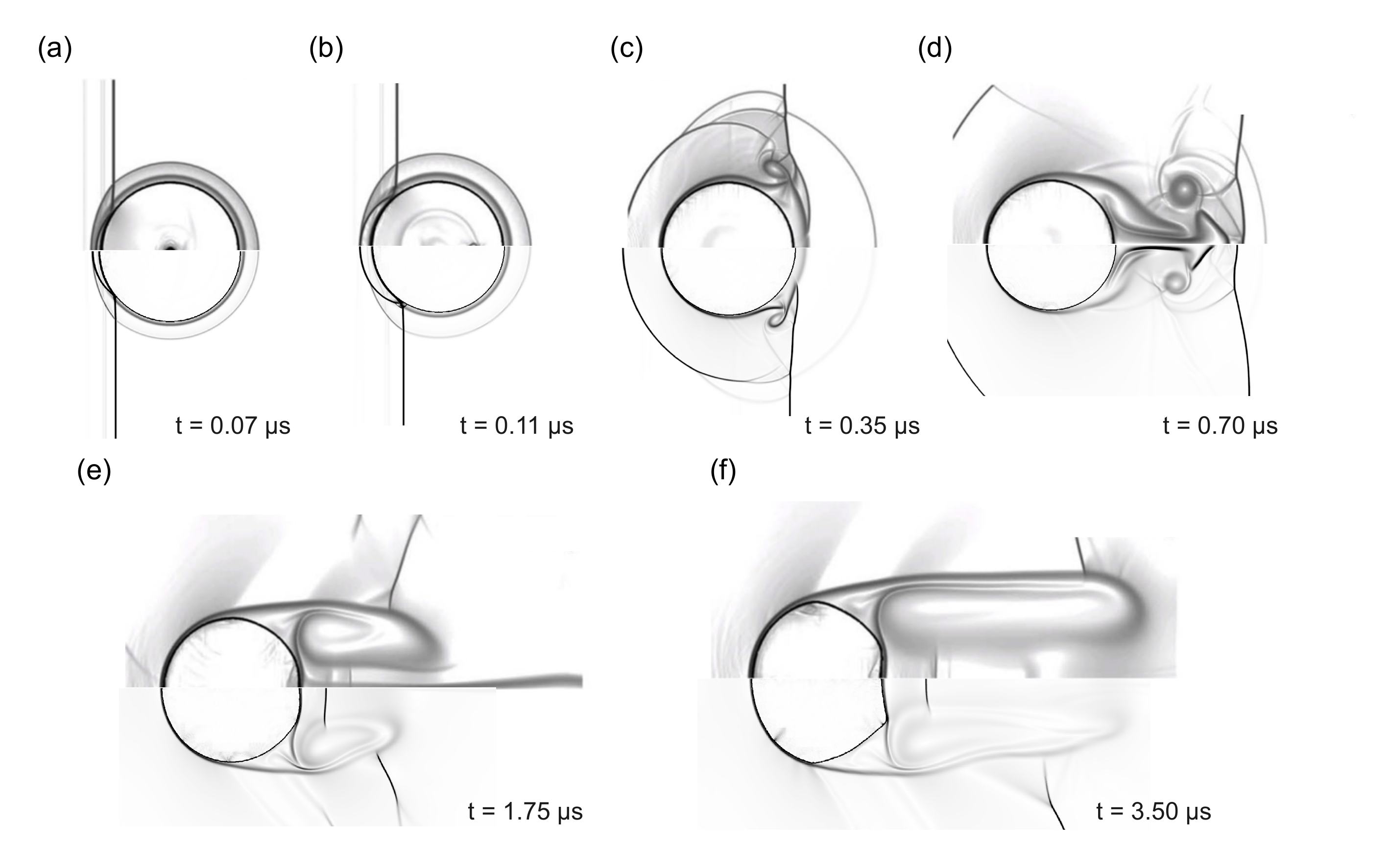}
  \caption{Numerical schlieren for a Mach 2 shock wave interacting with a vaporizing aluminum
  droplet. The upper half shows the benchmark result reprinted from Houim and Kuo, J. Comput. Phys. \textbf{235}, 865 (2013),
  with the permission of Elsevier. The lower half shows the present result.}
  \label{fig:shock-al-evap-schlieren}
\end{figure}

\begin{figure}[tbp]
  \centering
  \includegraphics[width=\linewidth]{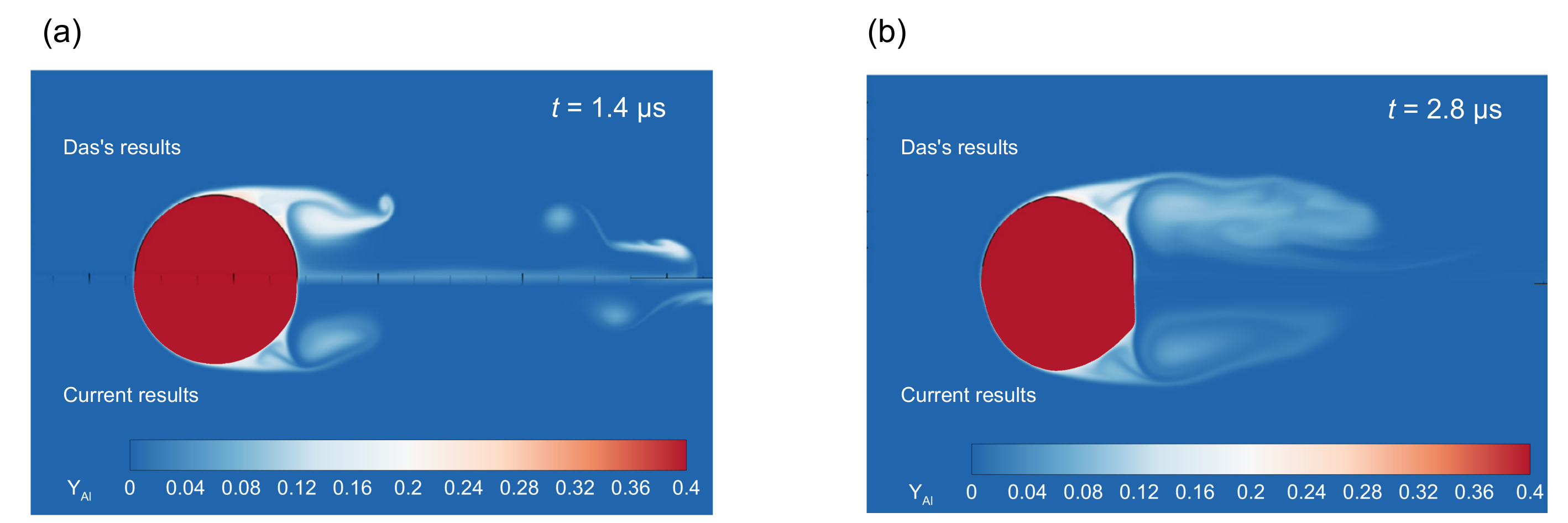}
  \caption{Mass fraction of Al vapor for a Mach 2 shock wave interacting with a vaporizing
  aluminum droplet. The upper half shows the benchmark result reprinted from Das and Udaykumar, J. Comput. Phys. \textbf{405},
  109005 (2020), with the permission of Elsevier. The lower half shows the present result.}
  \label{fig:shock-al-evap-vapor-fraction}
\end{figure}

\begin{figure}[tbp]
  \centering
  \includegraphics[width=0.4\linewidth]{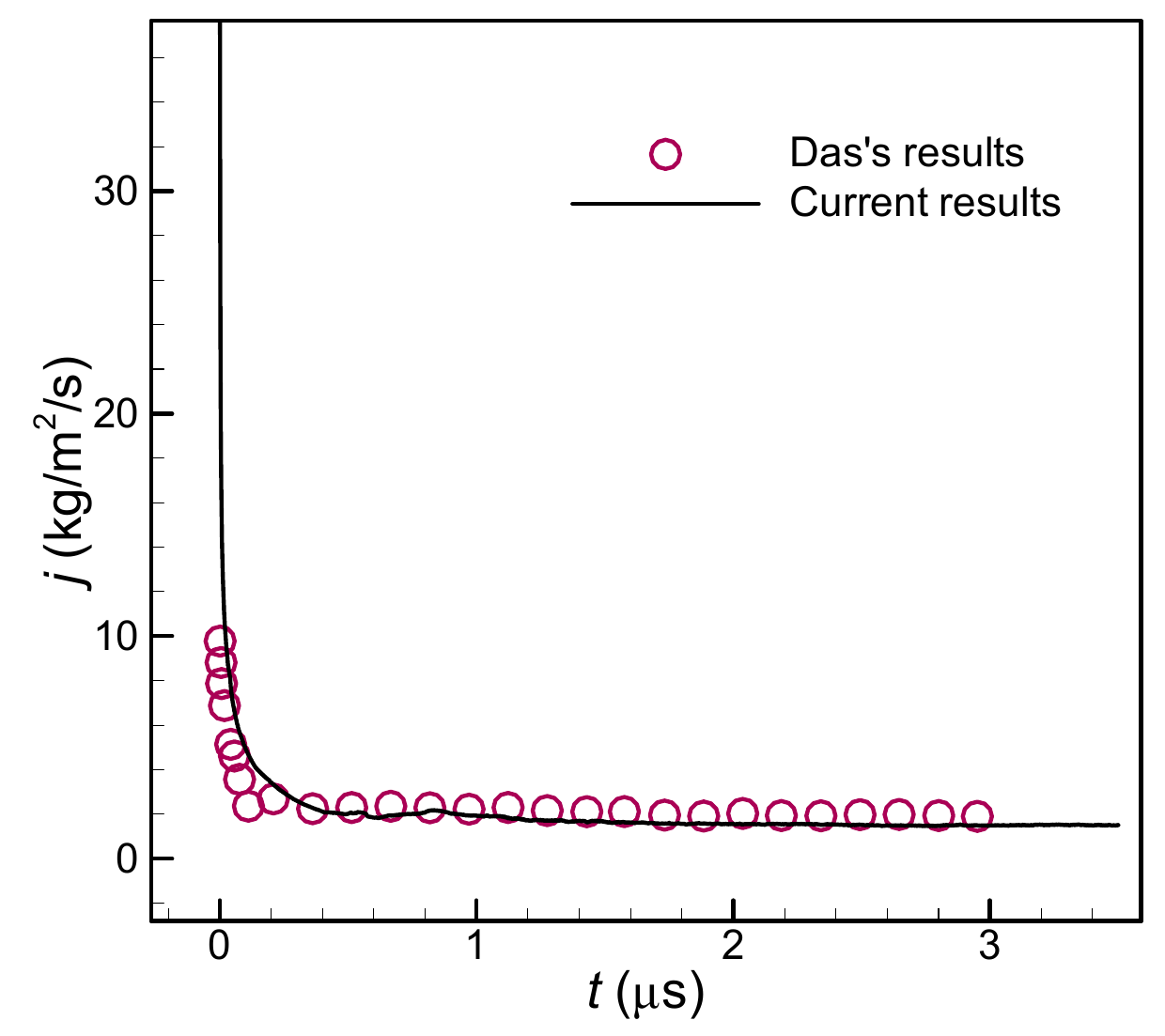}
  \caption{Average vaporization mass flux for a Mach 2 shock wave interacting with an
  vaporizing aluminum droplet: benchmark result of Das et
  al.~\cite{das2020shockVaporizationDroplets} and present study.}
  \label{fig:shock-al-evap-mass-flux}
\end{figure}

Fig.~\ref{fig:shock-al-evap-schlieren} compares the numerical schlieren fields with the
benchmark result of Houim and Kuo~\cite{houim2013ghostFluidPhaseChange}. The present simulation
reproduces the main wave structures observed in the benchmark, including the incident, reflected,
transmitted, and diffracted waves. It also captures the vaporization-induced shock waves
\cite{fechter2017sharpInterfacePhaseTransition,fechter2018riemannPhaseTransition} generated
before shock arrival, as shown in Figs.~\ref{fig:shock-al-evap-schlieren}(a) and (b).
The $Y_{\mathrm{Al}}$ distribution in Fig.~\ref{fig:shock-al-evap-vapor-fraction} also agrees
well with the benchmark result of Das and Udaykumar~\cite{das2020shockVaporizationDroplets}. After
shock impact, vapor generated near the droplet surface is stripped from the interface, convected
downstream, and transported into the recirculation region behind the droplet. The temporal
evolution of the average vaporization mass flux in
Fig.~\ref{fig:shock-al-evap-mass-flux} is in good agreement with the reference data.

\FloatBarrier
\subsubsection{Shock interaction with a condensing aluminum droplet}
\label{sec:2d-al-droplet-condensation}

We next consider shock interaction with a condensing aluminum droplet. Since no published
benchmark is available for this configuration, the case is used to demonstrate the behavior of
the present method for shock-driven condensation. The computational domain, Mach number, droplet
diameter, boundary conditions, and grid resolution are the same as those in
Sec.~\ref{sec:2d-al-droplet-evaporation}. Condensation is induced by initializing the gas mixture
in an aluminum-rich state, with
$X_{\mathrm{Al}}:X_{\mathrm{N_2}}:X_{\mathrm{O_2}}=1000:79:21$. All other initial conditions are
unchanged from the vaporization case.

\begin{figure}[!htbp]
  \centering
  \includegraphics[width=\linewidth]{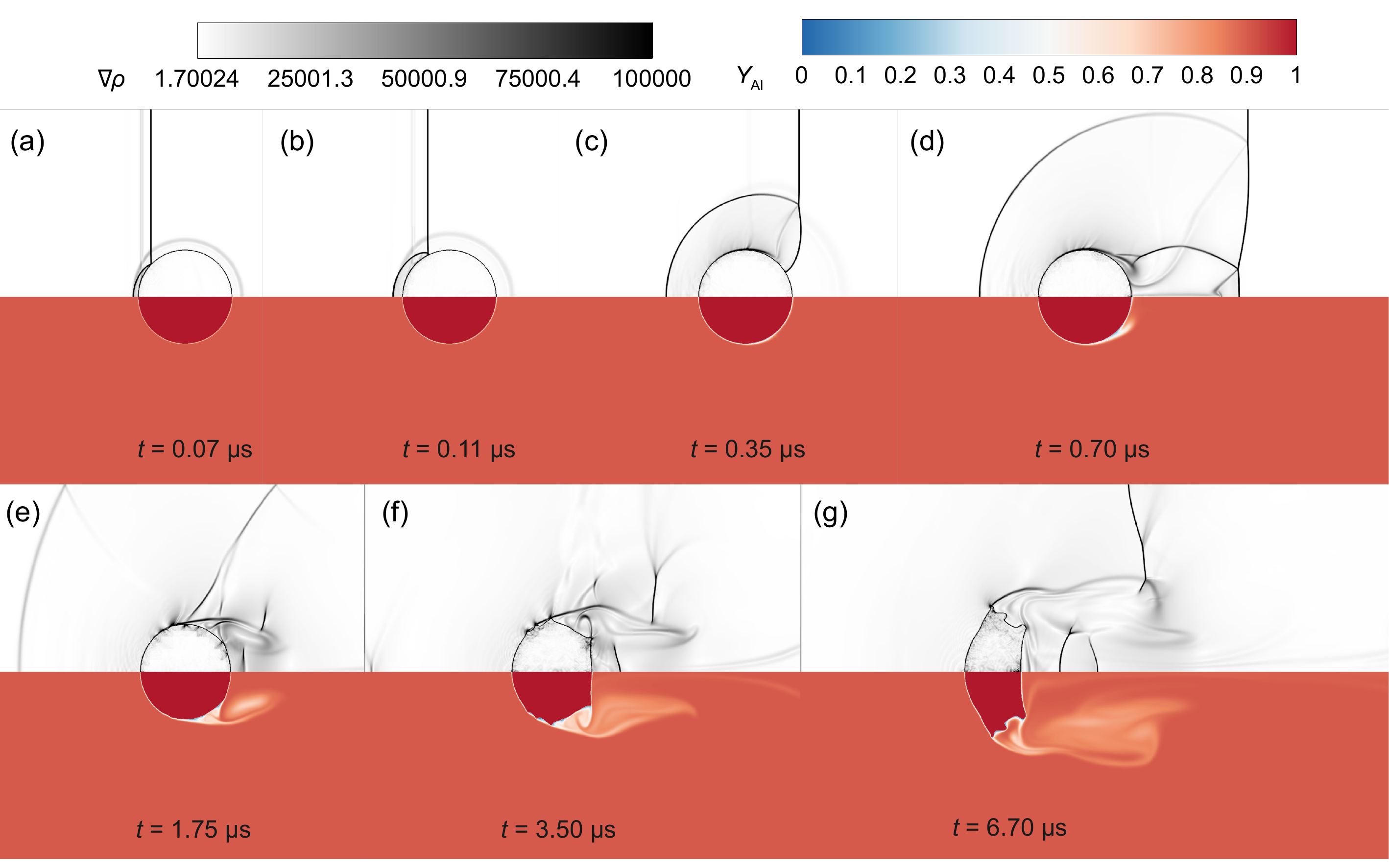}
  \caption{Time sequence of numerical schlieren (upper half) and the mass fraction of Al vapor
    (lower half) for a Mach 2 shock wave interacting with a condensing aluminum droplet.}
  \label{fig:shock-al-cond-schlieren-vapor}
\end{figure}

Fig.~\ref{fig:shock-al-cond-schlieren-vapor} shows the numerical schlieren and
$Y_{\mathrm{Al}}$ distributions. The schlieren contours contain the incident, reflected,
transmitted, and diffracted waves, as well as the downstream wake. Unlike the vaporization case,
no vaporization-induced shock waves appear near the droplet surface before shock impact. The
$Y_{\mathrm{Al}}$ distribution shows the opposite interfacial response: aluminum vapor is removed
from the gas side of the interface, forming a vapor-depleted layer that is stretched downstream by
the post-shock flow.

\begin{figure}[!htbp]
  \centering
  \includegraphics[width=0.65\linewidth]{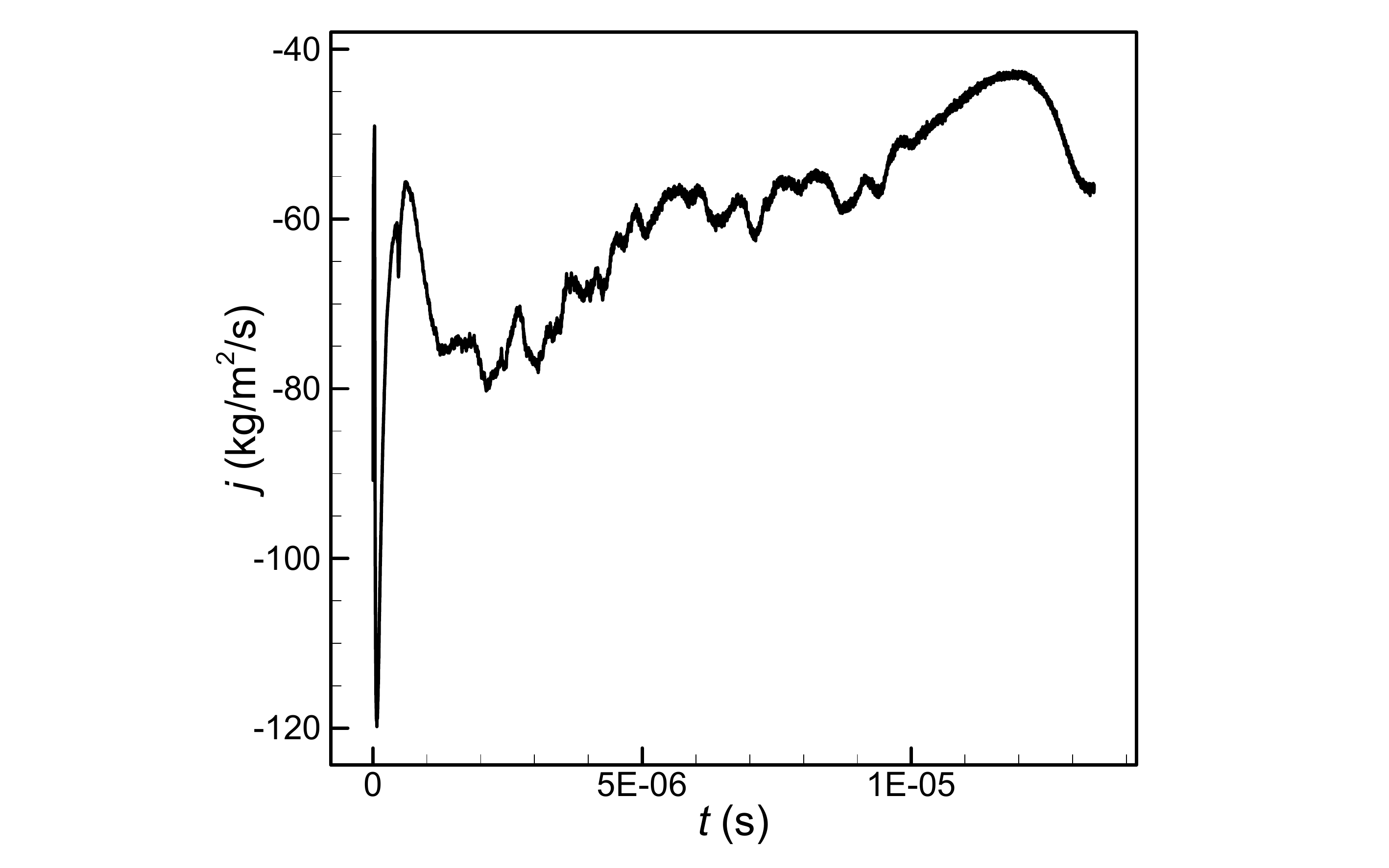}
  \caption{Temporal evolution of the average condensation mass flux for a Mach 2 shock wave
    interacting with a condensing aluminum droplet.}
  \label{fig:shock-al-cond-mass-flux}
\end{figure}

Fig.~\ref{fig:shock-al-cond-mass-flux} shows the temporal evolution of the average condensation
mass flux. The flux remains negative, confirming that the net phase change is directed from the
gas phase to the liquid phase. Its magnitude changes rapidly when the incident shock reaches the
droplet and then decreases gradually as the vapor-depleted layer and recirculating wake limit the
local supply of aluminum vapor to the interface.

\FloatBarrier
\subsubsection{Shock-induced aluminum droplet vaporization with chemical reactions}
\label{sec:2d-al-droplet-reaction}

\begin{figure}[!htbp]
  \centering
  \includegraphics[width=\linewidth]{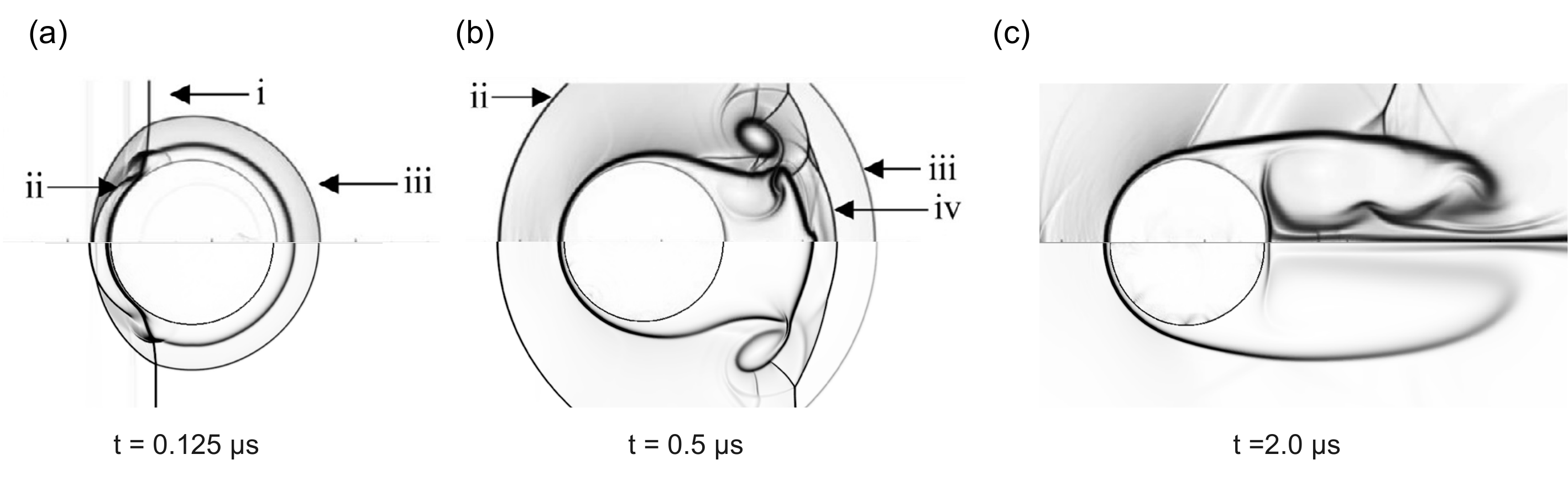}
  \caption{Numerical schlieren for a Mach 2 shock-droplet interaction with vaporization and
    chemical reactions. The upper half shows the benchmark result reprinted from Das and Udaykumar, Int. J. Multiphase Flow
    \textbf{134}, 103442 (2021), with the permission of Elsevier. The lower half shows the present result.}
  \label{fig:shock-al-react-schlieren}
\end{figure}

In this section, chemical reactions are added to the shock-induced vaporization problem. The initial
conditions are identical to those in Sec.~\ref{sec:2d-al-droplet-evaporation}, except that the
finite-rate aluminum oxidation mechanism is activated in the gas phase. The mechanism is the
same as that used in Sec.~\ref{sec:aluminum-slab}, following Houim and
Kuo~\cite{houim2013ghostFluidPhaseChange}.
Fig.~\ref{fig:shock-al-react-schlieren} shows that the numerical schlieren field reproduces the
benchmark wave structures~\cite{das2021reactingAluminumDroplets}. Compared with the
non-reacting case, gas-phase reactions thicken the aluminum-vapor layer and modify the
downstream flow. This enhancement is caused by aluminum-vapor consumption in the gas-side
boundary layer and heat release near the droplet, which strengthen the vaporization driving
force and increase the local gas temperature, respectively.

\begin{figure}[tbp]
  \centering
  \includegraphics[width=\linewidth]{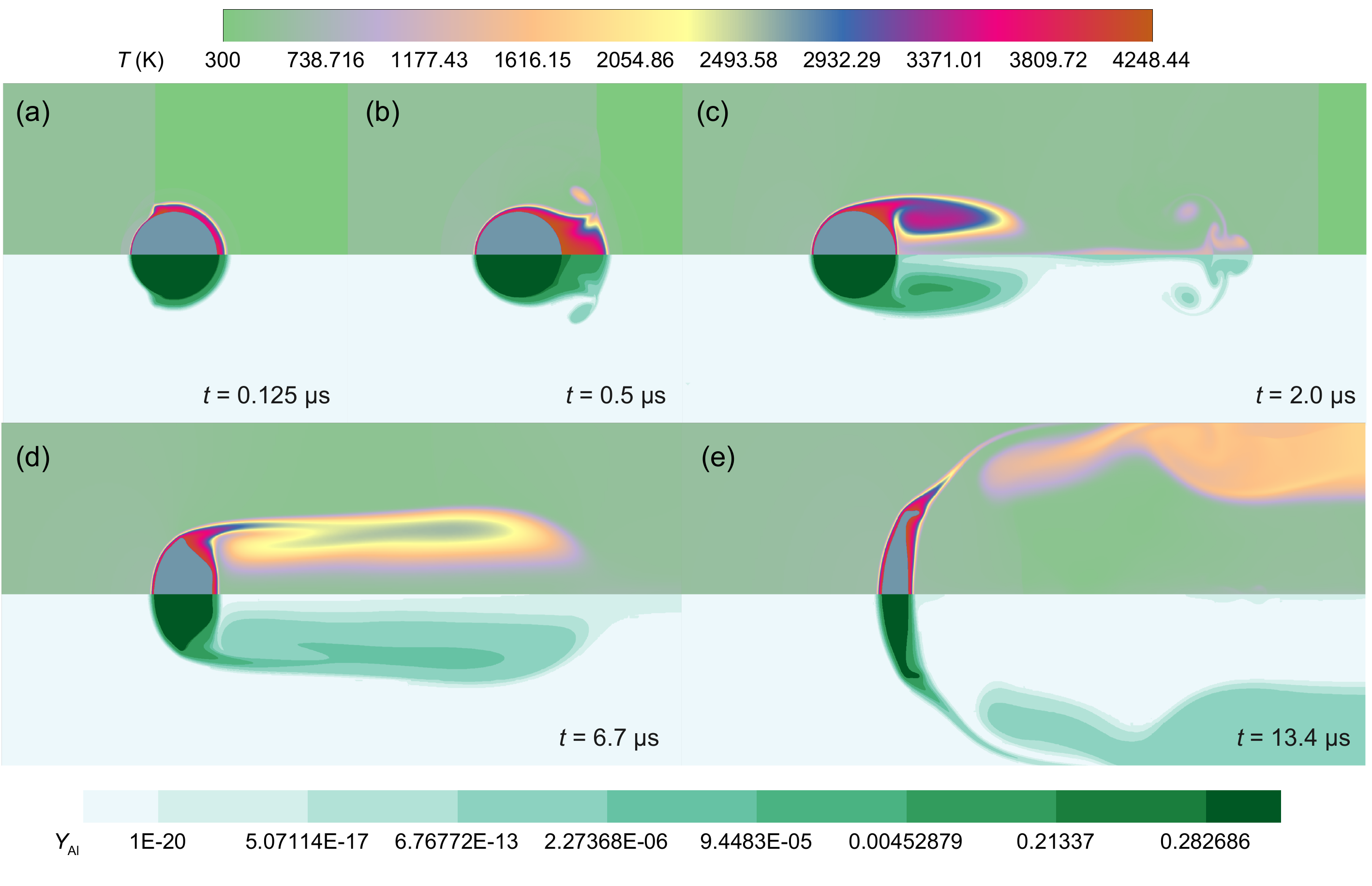}
  \caption{Time sequence of temperature (upper half) and the mass fraction of Al vapor (lower
    half) for a Mach 2 shock-droplet interaction with vaporization and chemical reactions.}
  \label{fig:shock-al-react-temperature-vapor}
\end{figure}

\begin{figure}[tbp]
  \centering
  \includegraphics[width=\linewidth]{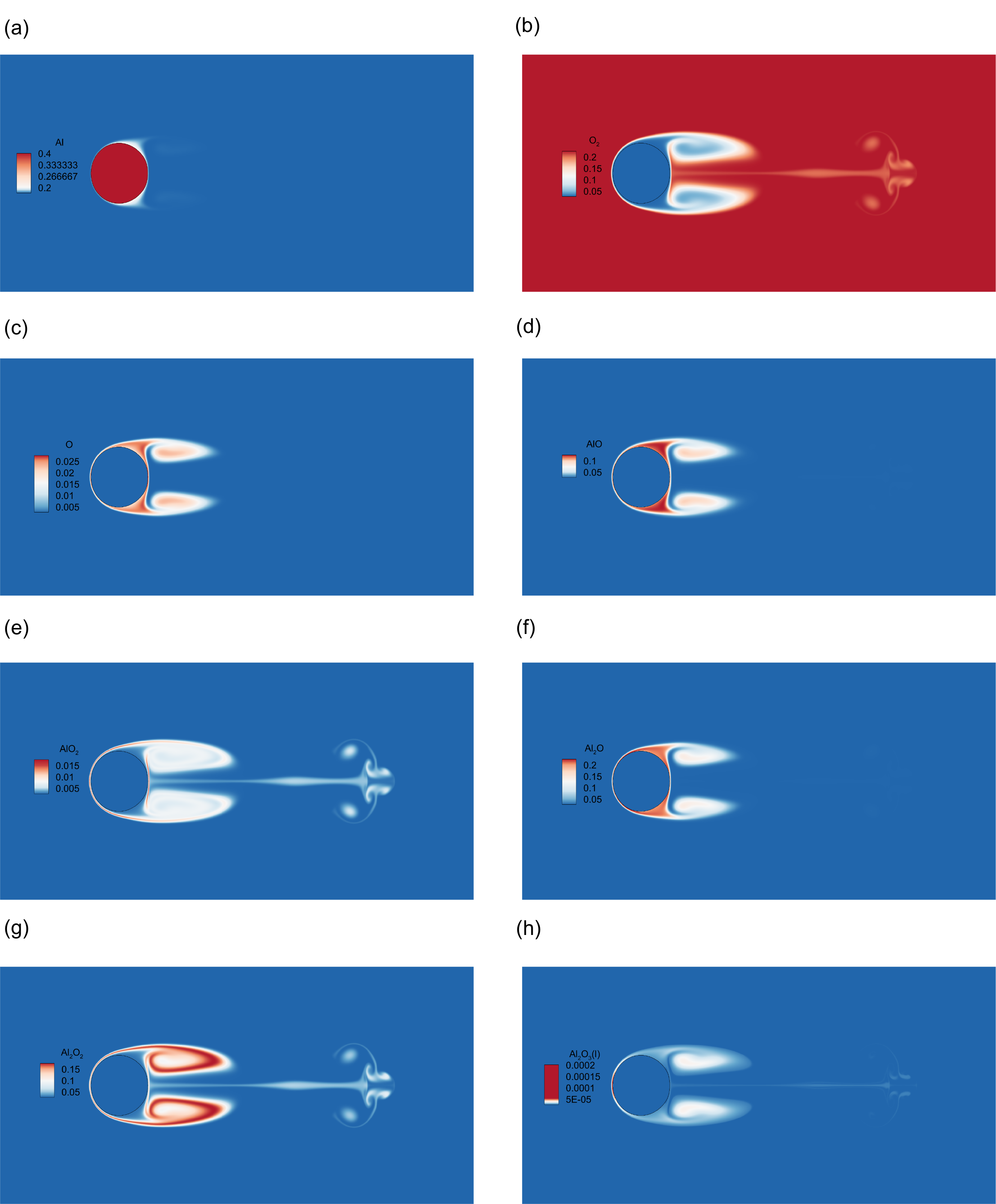}
  \caption{Computed species mass fractions for a Mach 2 shock-droplet interaction with
    vaporization and chemical reactions.}
  \label{fig:shock-al-react-species}
\end{figure}

The temperature and aluminum-vapor fields in
Fig.~\ref{fig:shock-al-react-temperature-vapor} further characterize the reacting shock-droplet
flow. In the non-reacting case, aluminum vapor is mainly stretched and convected downstream by the
post-shock flow. With reactions, aluminum oxidation consumes vapor and releases heat, thereby
enhancing vaporization and producing a high-temperature region around the droplet and in the
wake. The spatial correlation between this region and the vapor layer shows that the reaction is
localized where aluminum vapor mixes with the surrounding oxygen-rich gas. The species
distributions in Fig.~\ref{fig:shock-al-react-species} support this interpretation: oxygen is
depleted in the reaction layer, whereas aluminum oxide intermediates and products are
concentrated in the shear layer and wake, where aluminum vapor mixes with the surrounding
oxygen. The intermediate species peak mainly along the droplet boundary layer, and a small
amount of the $\mathrm{Al_2O_3(l)}$ reaction product accumulates near the windward stagnation
region. Compared with the non-reacting vaporization case in
Sec.~\ref{sec:2d-al-droplet-evaporation}, chemical reactions therefore provide an additional sink
for aluminum vapor and a localized heat source in the gas-side flow.

\begin{figure}[tbp]
  \centering
  \includegraphics[width=0.6\linewidth]{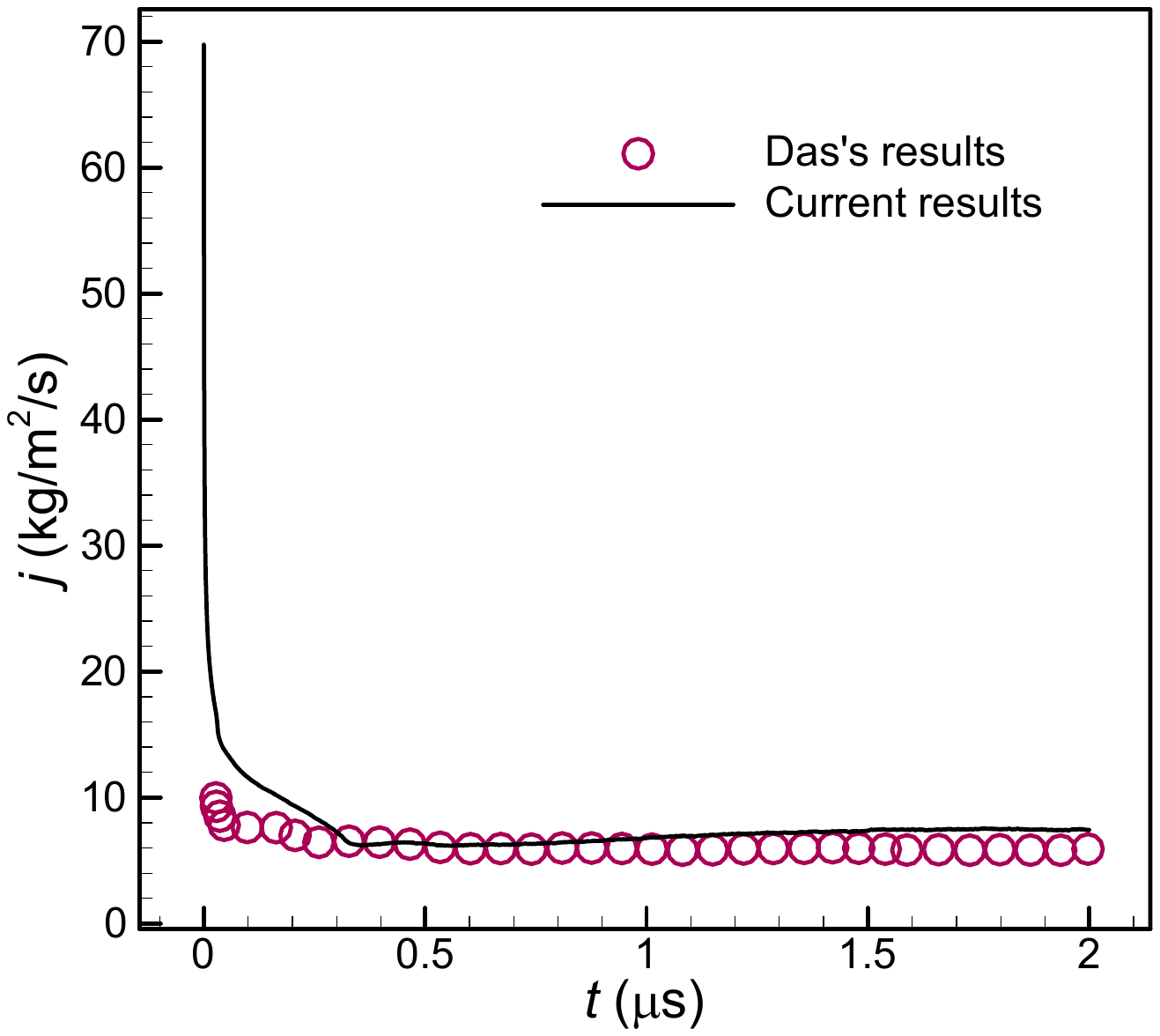}
  \caption{Average vaporization mass flux for a Mach 2 shock wave interacting with an
    vaporizing aluminum droplet with chemical reactions: benchmark result of Das and
    Udaykumar~\cite{das2021reactingAluminumDroplets} and present study.}
  \label{fig:shock-al-react-mass-flux}
\end{figure}

Consistent with these flow-field features, Fig.~\ref{fig:shock-al-react-mass-flux} compares the average vaporization
mass flux with the reacting-droplet benchmark. After the initial shock-induced transient, the
present result approaches the benchmark level and follows the same long-time trend. The reacting
case maintains a higher vaporization rate than the non-reacting case in
Fig.~\ref{fig:shock-al-evap-mass-flux}. This increase is attributed to aluminum-vapor consumption
in the gas-side reaction layer and chemical heat release, which strengthen the vaporization
driving force and raise the local gas temperature, respectively.

\FloatBarrier
\subsection{Detonation interaction with a water droplet}
\label{sec:2d-detonation-droplet}

We next consider the interaction between a gaseous detonation wave and a water droplet. Before
the two-phase calculation, a 1D hydrogen-oxygen detonation is simulated to verify the ability of
the present solver to reproduce the reference Zel'dovich--von Neumann--Doering (ZND) structure.
The gas mixture is initialized with the mole ratio
$X_{\mathrm{H_2}}:X_{\mathrm{O_2}}:X_{\mathrm{Ar}}=2:1:7$ at
$p_0=1.0\times10^5~\mathrm{Pa}$ and $T_0=293~\mathrm{K}$. A high-pressure,
high-temperature driver region with $p=1.0\times10^7~\mathrm{Pa}$,
$T=3000~\mathrm{K}$, and a length of $1.0~\mathrm{mm}$ is used to initiate the detonation.
The corresponding ZND solution obtained from SDToolbox
\cite{browne2018sdtoolbox} gives
$D_{\mathrm{CJ}}=1692.89~\mathrm{m/s}$,
$p_{\mathrm{VN}}=2.93~\mathrm{MPa}$,
$T_{\mathrm{VN}}=2049.9~\mathrm{K}$,
$p_{\mathrm{CJ}}=1.71~\mathrm{MPa}$, and
$T_{\mathrm{CJ}}=3074.3~\mathrm{K}$.
The hydrogen-oxygen reaction mechanism follows the PREMIX mechanism of
Kee et al.~\cite{kee1985premix}. It contains nine species,
$\mathrm{H_2}$, $\mathrm{O_2}$, $\mathrm{O}$, $\mathrm{H}$,
$\mathrm{OH}$, $\mathrm{HO_2}$, $\mathrm{H_2O_2}$,
$\mathrm{H_2O}$, and $\mathrm{Ar}$, and 18 reversible reactions.
The reaction rate follows the Arrhenius form of Eq.~\eqref{eq:eq21}. The complete reaction
mechanism and rate parameters are adopted from Kee et al.~\cite{kee1985premix}.

\begin{figure}[!htbp]
  \centering
  \includegraphics[width=\linewidth]{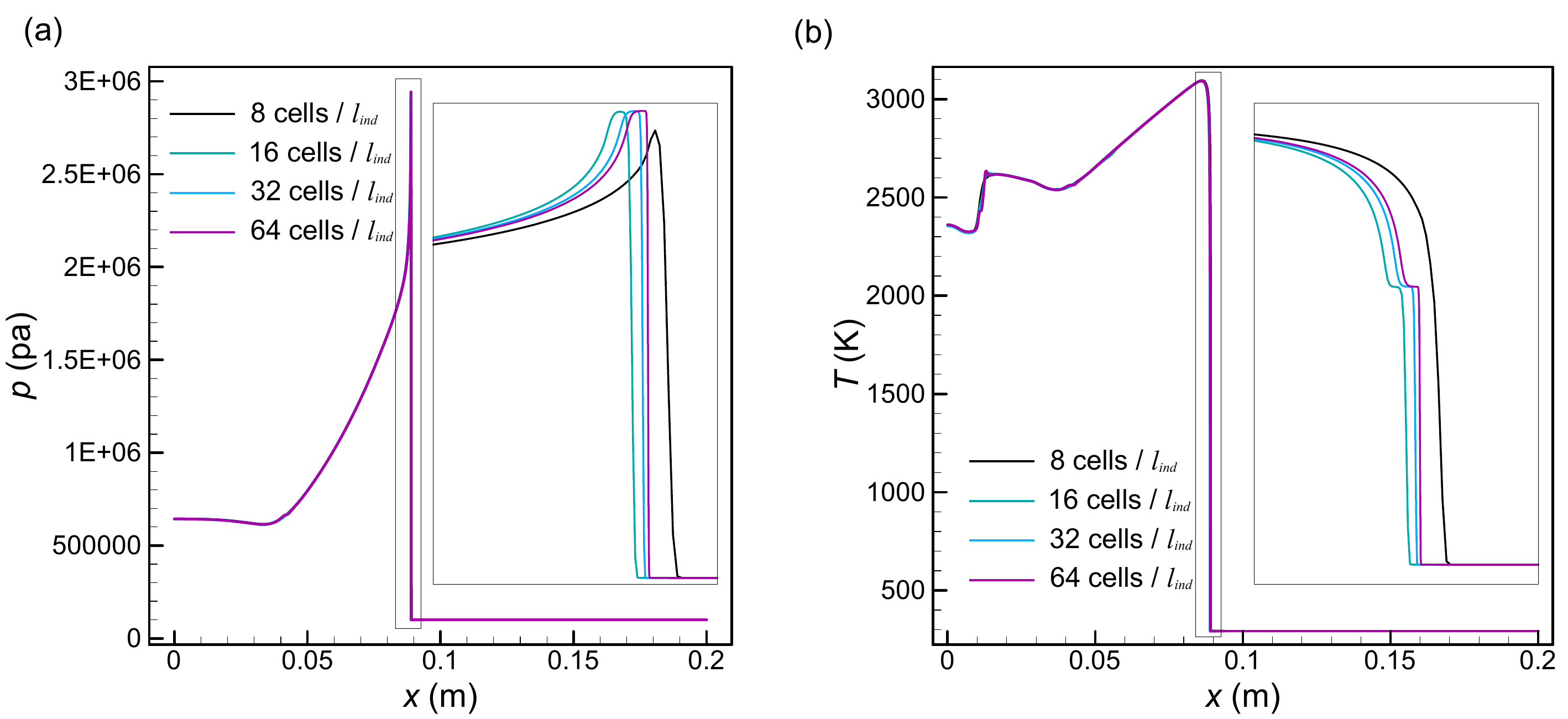}
  \caption{Verification of grid convergence for the 1D gaseous detonation. (a) Pressure and (b)
    temperature profiles.}
  \label{fig:detonation-grid-convergence}
\end{figure}

\begin{figure}[!htbp]
  \centering
  \includegraphics[width=\linewidth]{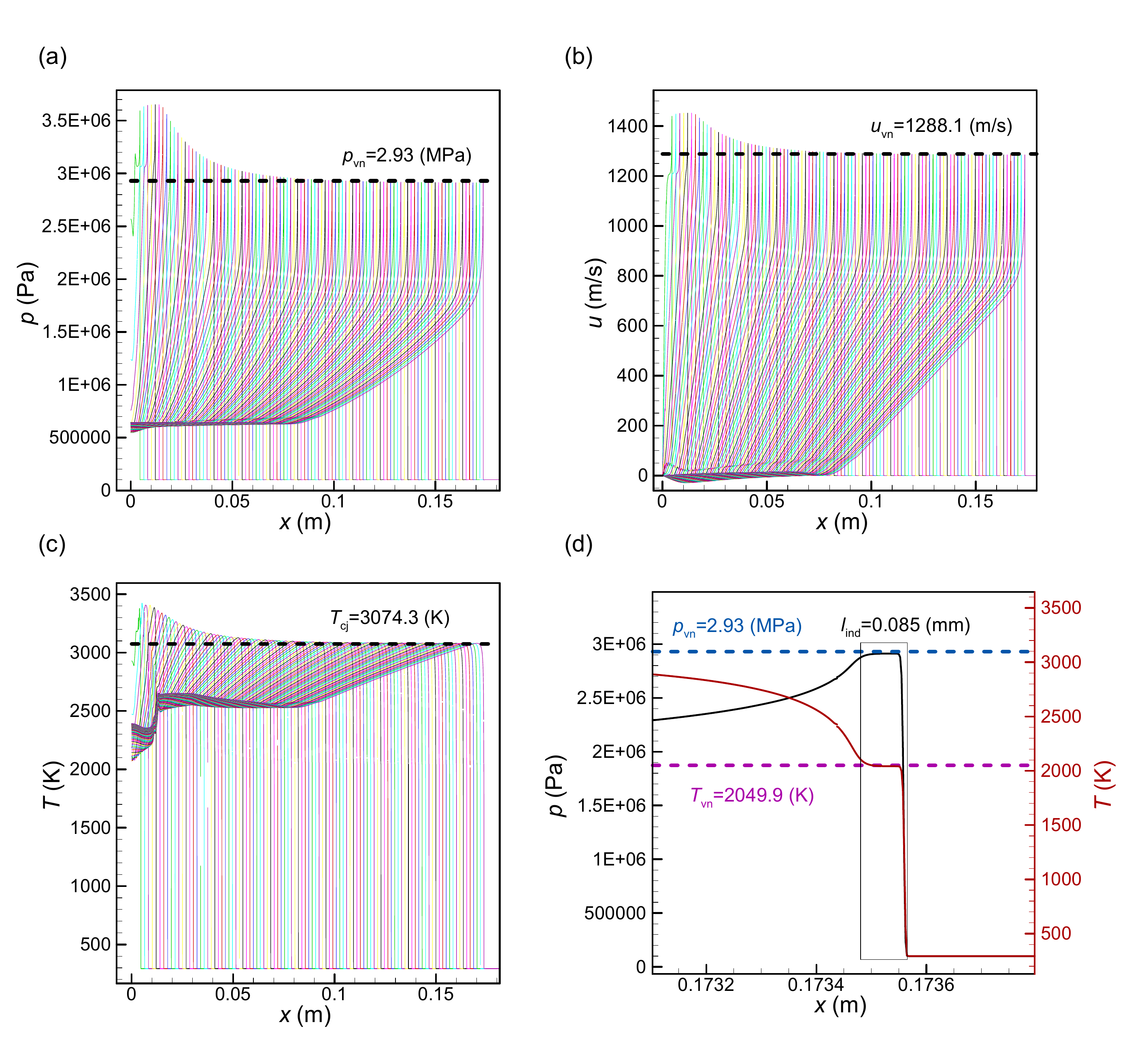}
  \caption{(a) Pressure, (b) velocity, and (c) temperature profiles at different time instants
    for the 1D gaseous detonation. The black dashed lines denote the reference solutions
    obtained from SDToolbox \cite{browne2018sdtoolbox}. (d) Comparison of the ZND structure
    between the present numerical results and SDToolbox.}
  \label{fig:one-dimensional-detonation}
\end{figure}

Fig.~\ref{fig:detonation-grid-convergence} shows the grid convergence of the 1D detonation. As
the grid is refined, the pressure and temperature plateaus behind the leading
shock are clearly resolved and remain unchanged upon further refinement.
Following Xu et al.~\cite{xu2024detonationWaterDroplet}, at least 10 grid cells are
required to adequately resolve the detonation induction zone. The resolved 1D detonation is then
compared with the ZND solution in Fig.~\ref{fig:one-dimensional-detonation}. The computed
pressure, velocity, and temperature profiles follow the ZND reference. In particular,
Fig.~\ref{fig:one-dimensional-detonation}(d)
shows that the induction length is accurately resolved and that the pressure and temperature in
the von Neumann region agree well with the ZND solution.

\begin{figure}[!htbp]
  \centering
  \includegraphics[width=\linewidth]{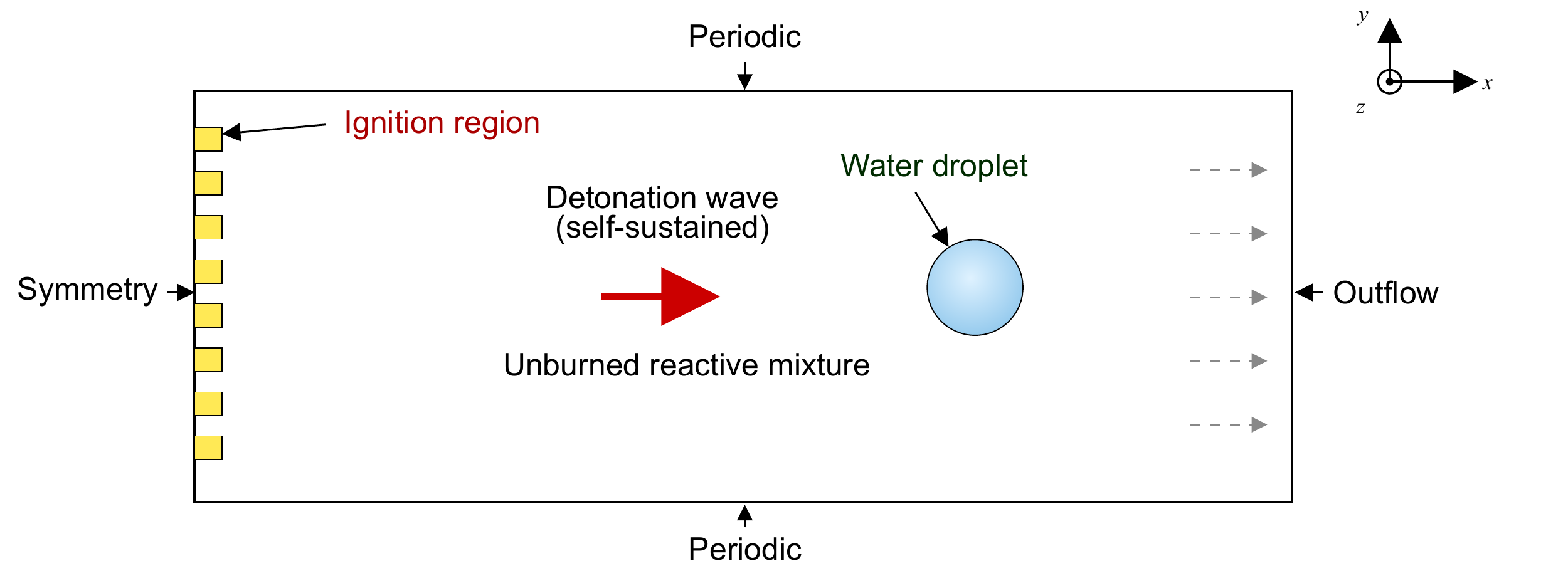}
  \caption{Schematic of the computational domain for the interaction between a detonation wave
    and a water droplet. The detonation wave is initiated by high-pressure and
    high-temperature square ignition regions, develops into a self-sustained wave, and
    subsequently interacts with the droplet.}
  \label{fig:detonation-droplet-domain}
\end{figure}

Having verified the 1D detonation structure, we next consider a two-dimensional
detonation-droplet interaction. The calculation follows the configuration of
Xu et al.~\cite{xu2024detonationWaterDroplet}. As shown in
Fig.~\ref{fig:detonation-droplet-domain}, the same initial gas mixture and ignition conditions
as the 1D detonation are used. The rectangular domain has a length of
$102.4~\mathrm{mm}$ and a height of $12.8~\mathrm{mm}$. Eight square ignition regions with a
side length of $1.0~\mathrm{mm}$ are placed near the left boundary to initiate the detonation.
A water droplet with diameter $D_0=3.2~\mathrm{mm}$ is centered at
$(x,y)=(80.0,6.4)~\mathrm{mm}$. The droplet is placed sufficiently downstream so that a stable
self-sustained detonation forms before reaching the droplet, as confirmed by the preceding 1D
calculation. The initial droplet velocity is zero, and the liquid state is
$p=1.01325\times10^5~\mathrm{Pa}$ and $T=365~\mathrm{K}$. Periodic conditions are imposed at the
transverse boundaries, symmetry is imposed at the left boundary, and an outflow condition is used
at the right boundary. The effective resolution is $16384\times2048$, corresponding to 2048 grid
cells along the short side of the domain. Based on the ZND induction length of
$0.085~\mathrm{mm}$, the induction zone is resolved by approximately 14 grid cells, exceeding the
minimum of 10 cells recommended by
Xu et al.~\cite{xu2024detonationWaterDroplet}.

\begin{figure}[!htbp]
  \centering
  \includegraphics[width=0.7\linewidth]{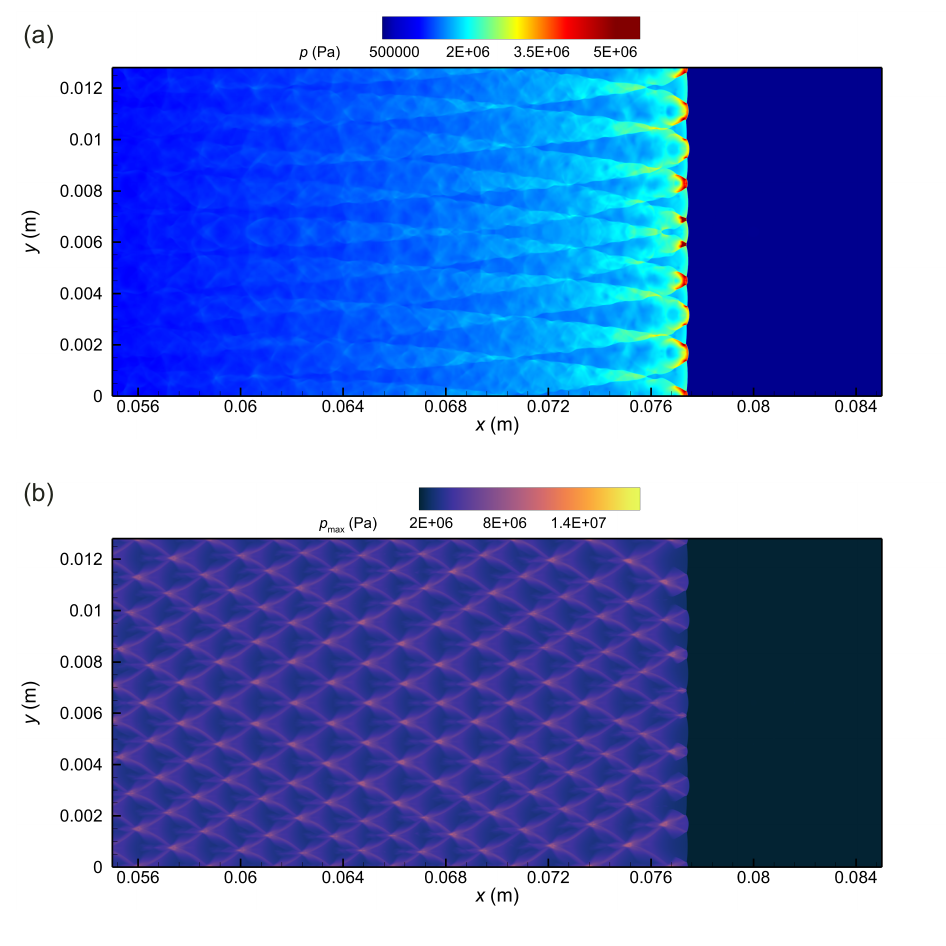}
  \caption{Flow fields of the 2D gaseous cellular detonation wave: (a) pressure and (b)
    numerical smoked foil.}
  \label{fig:cellular-detonation-pressure}
\end{figure}

\begin{figure}[!htbp]
  \centering
  \includegraphics[width=\linewidth]{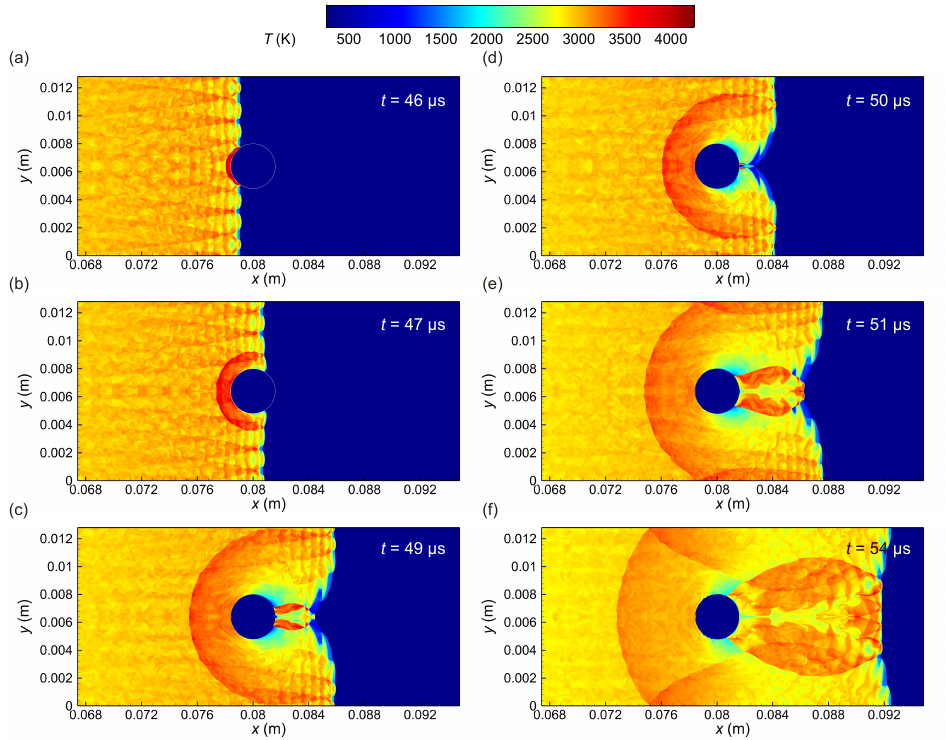}
  \caption{Numerical temperature contours for the interaction between a detonation wave and a
    water droplet at different time instants. The white lines denote the deformed water droplet
    interface.}
  \label{fig:detonation-droplet-temperature}
\end{figure}

Fig.~\ref{fig:cellular-detonation-pressure} shows the pressure field and the numerical smoked
foil before the detonation reaches the droplet. The detonation front exhibits a cellular
structure, with transverse waves and triple-point tracks forming a regular smoked-foil pattern.
This confirms that the detonation has developed from the initial ignition blocks into a
self-sustained cellular wave before interacting with the water droplet.
The subsequent interaction with the water droplet is shown in
Fig.~\ref{fig:detonation-droplet-temperature}. When the detonation front reaches the droplet, it
is reflected from the windward surface, producing a high-temperature compressed region in the
burned gas. The reflection first appears as a regular reflection and then evolves into a Mach
reflection as the interaction proceeds. When the cellular detonation diffracts around the
leeward side of the droplet, the leading shock is weakened and becomes locally decoupled from
the heat-release zone, causing a transient extinction behind the droplet. The extinguished waves
from the upper and lower sides subsequently merge near the rear pole, where the temperature
rises sharply and local re-ignition occurs. The re-ignited wave then propagates outward as an
overdriven detonation and reconnects with the weakened cellular front downstream. These features
are consistent with the detonation-droplet interaction reported by
Xu et al.~\cite{xu2024detonationWaterDroplet}, indicating that the present method captures the
coupled shock reflection, diffraction, local extinction, and re-ignition processes associated
with detonation interaction with an evaporating droplet.

\FloatBarrier
\subsection{3D shock interaction with a vaporizing aluminum droplet}
\label{sec:three-dimensional}

We finally extend the Mach 2 shock interaction with a vaporizing aluminum droplet to 3D. The
initial conditions follow the 2D axisymmetric case in
Sec.~\ref{sec:2d-al-droplet-evaporation}. A quarter-symmetric domain
is used, with two symmetry planes passing through the droplet center. The outer transverse and
downstream boundaries are treated as outflow boundaries, whereas the upstream boundary is
prescribed as an inflow boundary. A five-level MR adaptive grid is used, giving an effective
resolution of $512\times512\times2048$. This corresponds to 256 grid cells across the initial
droplet diameter.

\begin{figure}[!htbp]
  \centering
  \includegraphics[width=\linewidth]{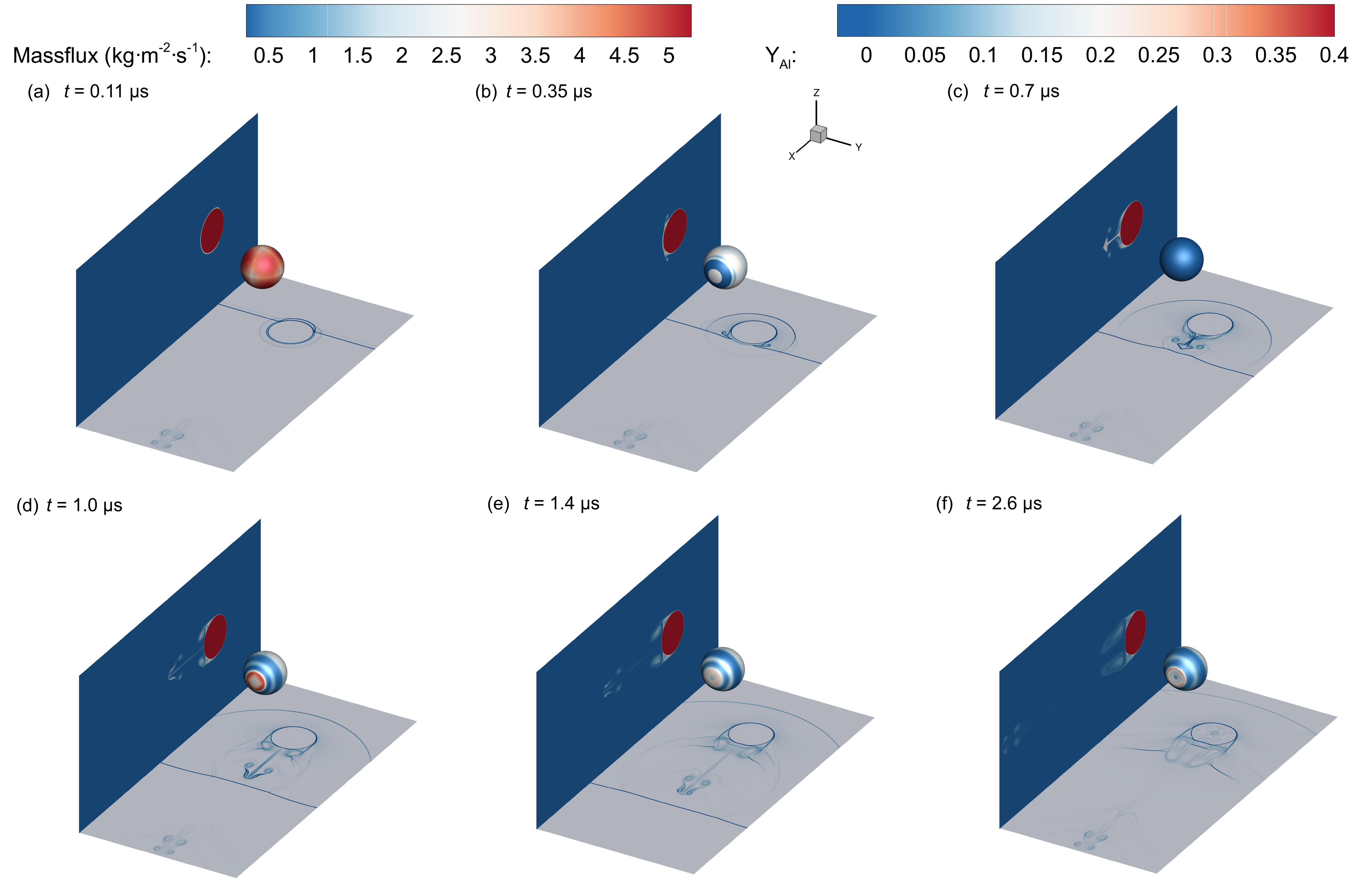}
  \caption{3D shock interaction with a vaporizing aluminum droplet. The droplet
    surface is colored by the vaporization mass flux, and the cut planes show the aluminum-vapor
    mass fraction and wave structures.}
  \label{fig:three-dimensional-shock-al}
\end{figure}

Fig.~\ref{fig:three-dimensional-shock-al} shows the surface vaporization mass flux on the
$\phi=0$ droplet interface and the aluminum-vapor mass fraction on representative cut planes.
Before shock arrival, the large mass flux indicates strong vaporization from the heated droplet.
After shock impingement, the flux decreases over most of the interface, consistent with the
suppressed vaporization observed in the 2D axisymmetric case. The remaining nonuniform flux and
the aluminum vapor are then transported toward the lateral and downstream wake regions. This 3D
extension retains the principal shock-interface and vapor-transport features while demonstrating
that the present conservative sharp-interface and diffuse-interface coupling method can handle a fully three-dimensional moving
interface with phase change and multi-species transport.

\FloatBarrier
\section{Conclusions}
\label{sec:conclusions}

A conservative sharp-interface and diffuse-interface coupling method has been developed for
compressible two-phase multi-species flows with phase change and chemical reactions. The
liquid--gas interface is represented sharply, whereas gas-phase species transport and chemical
reactions are handled by a diffuse-interface method.
These two descriptions are coupled conservatively through the interfacial flux treatment.
Compared with GFM-based methods
for multi-species phase change
\cite{houim2013ghostFluidPhaseChange,das2020shockVaporizationDroplets,
das2021reactingAluminumDroplets}, the present method couples the phase-changing liquid and gas
phases through interfacial fluxes obtained from a multi-species phase-change Riemann problem, so
that strict conservation of mass, momentum, and energy is ensured. Compared with conservative
sharp-interface methods developed for pure liquid-vapor systems
\cite{lauer2012cavitationBubbleDynamics,paula2019waterDropExplosion,
long2023conservativePhaseChange}, the present method extends the
interfacial Riemann coupling to gas mixtures containing both condensable vapor and
non-condensable species, and incorporates species diffusion and chemical reactions. The
interfacial energy jump condition is modified by evaluating the gas-side phase-change energy
flux with the internal energy of the phase-changing vapor species rather than that of the gas
mixture. The impulsive
condensation test shows that this species-selective energy jump removes the spurious pressure and
temperature structures obtained when the mixture internal energy is used. The present method
has been assessed using test cases involving evaporation, condensation, shock-droplet interaction
with vaporizing, condensing, and reacting aluminum droplets, and detonation-droplet interaction.
The results agree with reference solutions and available benchmark data, confirm strict
conservation, and demonstrate the ability of the method to treat
strongly compressible wave-interface interactions with phase change, species transport, and
chemical reactions. In the present method, the liquid phase is treated as a pure liquid composed
of the phase-changing species. In practical applications, however, the liquid phase may also
consist of multiple species. Extension to liquid mixtures in which only one component undergoes
phase change is a natural next step for the present method. When multiple liquid species
can undergo phase change, however, additional interfacial thermodynamic closure will be required
and will be considered in future work.

\section*{CRediT authorship contribution statement}

\textbf{Jiaxi Song:} Conceptualization, Methodology, Software, Validation, Investigation,
Visualization, Writing -- original draft. \textbf{Yunzhang Tian:} Data curation.
\textbf{Shucheng Pan:} Conceptualization, Methodology,
Supervision, Funding acquisition, Writing -- review \& editing.

\section*{Declaration of competing interest}

The authors declare that they have no known competing financial interests or personal
relationships that could have appeared to influence the work reported in this paper.

\section*{Acknowledgements}

This work was supported by the National Natural Science Foundation of China (Grants
Nos.~1247021811 and 11902271). The authors thank Yixuan Lian (Northwestern Polytechnical
University) for helpful discussions on numerical methods for multi-species flows and chemical
reactions.

\section*{Data Availability}

Data will be made available on request.

\bibliographystyle{elsarticle-num}
\bibliography{references}

\end{document}